\documentclass[%
reprint,
amsmath,amssymb,
aps,
pre,
]{revtex4-2}

\usepackage{graphicx}
\usepackage{dcolumn}
\usepackage{bm}
\usepackage[colorlinks=true]{hyperref}
\usepackage{mwe}
\usepackage{color}
\usepackage[dvipsnames]{xcolor}

\newcommand{\blue}[1]{\textcolor{black}{#1}}

\begin{document}
	
	\title {Energy and information flows in autonomous systems} 
	
	\author{Jannik Ehrich}
	\email{jehrich@sfu.ca}
	
	\author{David A.\ Sivak}
	\email{dsivak@sfu.ca}
	\address{Department of Physics, Simon Fraser University, Burnaby, BC, V5A 1S6 Canada}
	
	\date{\today}
	
	\begin{abstract}
		Multi-component molecular machines are ubiquitous in biology. We review recent progress on describing their thermodynamic properties using autonomous bipartite Markovian dynamics. The first and second laws can be split into separate versions applicable to each subsystem of a two-component system, illustrating that one can not only resolve energy flows between the subsystems but also information flows quantifying how each subsystem's dynamics influence the joint system's entropy balance. Applying the framework to molecular-scale sensors allows one to derive tighter bounds on their energy requirement. Two-component strongly coupled machines can be studied from a unifying perspective quantifying to what extent they operate conventionally by transducing power or like an information engine by generating information flow to rectify thermal fluctuations into output power.
	\end{abstract}
	
	\maketitle
	
	\section{Introduction}
	Livings things are fundamentally out of thermodynamic equilibrium~\cite{Schrodinger:WhatIsLife}. Maintaining this state requires a constant flow of energy into them accompanied by dissipation of heat into their environment. Quantifying these flows is straightforward for macroscopic systems but much less so on the small scales of molecular machinery. The advent of ever-more-refined experimental equipment capable of probing small-scale thermodynamics has led to the burgeoning field of stochastic thermodynamics~\cite{Seifert2012_Stochastic,Jarzynski2011_Equalities,VandenBroeck2015_Ensemble,Peliti2021_book}. Within this theory, energy flows are deduced from the thermally influenced stochastic dynamics of small-scale systems, permitting quantification of heat dissipation and energetic requirements of diverse experimental setups as well as molecular biological machinery.

	\subsection{Information thermodynamics}
	Information plays an interesting and, at times, adversarial role in thermodynamics. At the dawn of statistical mechanics, Maxwell illustrated the counterintuitive role of information by arguing that an intelligent \emph{demon} could separate gas molecules according to their velocity with seemingly no expense of energy, apparently contradicting the second law~\cite{Maxwell1867_Life}. Resolving this paradox plagued physicists for a century~\cite{Maxwell1867_Life,Leff2003_Maxwells}, leading to well-known contributions from Szilard~\cite{Szilard1929}, Landauer~\cite{Landauer1961_Irreversibility}, and Bennett~\cite{Bennett1982_The_thermodynamics} ultimately showing that acquiring, processing, and storing information incurs thermodynamic costs that balance or exceed any benefit gained from it.
	
	Within the theory of stochastic thermodynamics, information has been incorporated in various ways, including measurement and feedback~\cite{Cao2009_Thermodynamics,Sagawa2010_Generalized,Ponmurugan2010_Generalized,Horowitz2010_Nonequilirbium,Sagawa2012_Nonequilibrium,Sagawa2012_Fluctuation,Sagawa2014_Role} performed by an experimenter on a system, and a system interacting with information reservoirs~\cite{Mandal2012_solvable_model,Mandal2013_Refrigerator,Barato2014_Unifying,Barato2014_Information_Reservoirs}; this established information as a proper thermodynamic resource~\cite{Parrondo2015_Thermodynamics} that sets limits on system capabilities similar to work and free energy. Diverse theoretical works~\cite{Bauer2012_Efficiency,Schmitt2015_Molecular,Bechhoefer2015,Still2020_cost_benefit_memory,Lucero2021_Maximal,Ehrich2022_Energetic,Still2021_Partially} and experimental realizations~\cite{Serreli2007_Molecular,Bannerman2009_Single-photon,Toyabe2010_Experimental,Koski2014_Experimental_Obs,Koski2014_Experimental_Realiz,Koski2015_On-chip,Vidrighin2016_Photonic,Camati2016_Experimental,Chida2017_Power,Cottet2017_Observing,Paneru2018_Losless,Masuyama2018_Information-to-work,Naghiloo2018_Information,Admon2018_Experimental,Paneru2018_Optimal,Ribezzi2019_Large,Paneru2020_Efficiency,Saha2021_Maximizing,Saha2022_Bayesian} illustrate information-powered engines.

	\subsection{Autonomous and complex systems consisting of subsystems}
	Small-scale information thermodynamics is also relevant for biological systems such as molecular machines and molecular-scale sensors~\blue{\cite{Amano2022_Insights,Takaki2022_Information}}. Understanding living systems at small scales and advancing the design of nanotechnology~\blue{\cite{Wilson2016_An_autonomous}} requires extending thermodynamics beyond conventional contexts: Instead of the scripted experimental manipulation of time-dependent control parameters, living systems are autonomous, driven out of equilibrium by steady-state nonequilibrium boundary conditions.
	
	Moreover, embracing more of the complexity of biology, we seek understanding beyond the interactions of a system with weakly coupled baths, to encompass interactions among strongly coupled subsystems
	\blue{~\cite{Feniouk1999_ATP-synthase,Toyabe2011_Thermodynamic,Lathouwers2020_Nonequilibrium,Large2021_Free-energy}.}
	Lacking a clear separation between a measurement that collects information about a system and feedback that acts on this information
	\blue{~\cite{Shiraishi2015_Role},}
	in autonomous systems it is more practical to differentiate between subsystems: An \emph{upstream} system that generates information for a \emph{downstream} system to react to or exploit.
	
	While in the non-autonomous setup apparent second-law violations result from not correctly accounting for non-autonomous interventions by an experimenter
	\blue{~\cite{Sagawa2010_Generalized,Ponmurugan2010_Generalized,Horowitz2010_Nonequilirbium,Sagawa2012_Nonequilibrium,Sagawa2012_Fluctuation,Sagawa2014_Role,Ehrich2022_Energetic},}
	in autonomous multi-component systems they can be traced back to thermodynamic accounting that ignores the strong coupling
	\blue{~\cite{Horowitz2014_Thermodynamics,Horowitz2015_Multipartite,Shiraishi2015_Role,Freitas2021_Characterizing}.}
	
	In its simplest form such an autonomous setup is realized by a downstream molecular sensor that reacts to an independent upstream signal
	\blue{~\cite{Barkai1997_Robustness,Sourjik2012_Responding,Mehta2012_Energetic,Barato2013_Rate,Barato2013_Information-theoretic,Barato2014_Efficiency}.}
	More complex interactions are realized by two-component strongly coupled molecular machines in which the dynamics of each component is influenced by the other
	\blue{~\cite{Junge2015_ATP_Synthase,Lathouwers2020_Nonequilibrium,Lathouwers2022_Internal}}
	and by assemblies of molecular transport motors that collectively pull cargo
	\blue{~\cite{Leopold1992_Association,Rastogi2016_Maximum}.}
	
	Here we focus on such autonomous systems, collecting results that extend information thermodynamics to contexts lacking explicit external measurement and feedback, and showcase that bipartite Markovian dynamics and information flow are versatile tools to understand the thermodynamics and performance limits of these systems.

	\subsection{Objectives and organization}
	Our aims with this review are to:
	\begin{enumerate}
		\item Build on stochastic thermodynamics to give a gentle introduction to the information-flow formalism, deriving all necessary equalities and inequalities and relating the different names and concepts for similar quantities that appear throughout the literature. Section~\ref{sec:bipartite_dyn} introduces bipartite dynamics and establishes the notation. Sections~\ref{sec:energy_flows} and \ref{sec:entropy_balance} deal with energy and information flows in these bipartite systems in general, while section~\ref{sec:Nostalgia} compares various similar information-flow measures.
		\item Collect results valid for biomolecular sensors, for which the information-flow formalism produces a tighter second law. These are contained in section~\ref{sec:sensor_setups}.
		\item Address engine setups and show that the information-flow formalism advances understanding of autonomous two-component engines simultaneously as work and information transducers (section~\ref{sec:engine_setups}).
	\end{enumerate}

	\subsection{Related reviews}
	The information-flow formalism is firmly rooted in the theory of stochastic thermodynamics. Recent reviews include a comprehensive one by Seifert~\cite{Seifert2012_Stochastic} and reviews by Jarzynski~\cite{Jarzynski2011_Equalities} (focusing on nonequilibrium work relations), Van den Broeck and Esposito~\cite{VandenBroeck2015_Ensemble} (explicitly dealing with jump processes), and Ciliberto~\cite{Ciliberto2017_Experiments} (on experiments in stochastic thermodynamics). The recent book by Peliti and Pigolotti~\cite{Peliti2021_book} also gives a pedagogical introduction to the field.
	Information thermodynamics itself has recently been reviewed by Parrondo \emph{et al.}~\cite{Parrondo2015_Thermodynamics}.
	
	Turning to molecular machinery, the working principles of Brownian motors have been reviewed by Reimann~\cite{Reimann2002_Brownian}. General aspects of molecular motors can be found in the reviews by Chowdhury~\cite{Chowdhury2013_Stochastic} and Kolomeisky~\cite{Kolomeisky2013_Motor}. Brown and Sivak~\cite{Brown2020_Theory} focus on the transduction of free energy by nanomachines, while reviews by Silverstein~\cite{Silverstein2014_Exploration} and Li and Toyabe~\cite{Li2020_Efficiencies} specifically treat the efficiencies of molecular motors.

	\section{Bipartite dynamics} \label{sec:bipartite_dyn}
	We consider a mesoscopic composite system whose state at time $t$ is denoted by $z(t)$. Due to thermal fluctuations, its dynamics are described by a Markovian stochastic process defined by a \emph{Master equation}~\cite{vanKampen2007_Stochastic,Gardiner2004_Handbook}:
	\begin{equation}
		\frac{\mathrm d}{\mathrm dt}p_t(z) = \sum_{z'} \left[ R(z|z';t) \, p_t(z') - R(z'|z;t) \, p_t(z) \right]\,, \label{eq:master_equation}
	\end{equation}
	where
	\blue{$p_t(z)$ is the probability to find the composite system in state $z$ at time $t$ and}
	the transition rates $R(z|z';t)$ (sometimes also called the \emph{generator}) encode the jump rates from state $z'$ to state $z$. For convenience, we assume a discrete state space; however, all results can easily be translated into continuous state-space dynamics, as we allude to in section~\ref{sec:continuous_state-spaces}.
	\blue{If multiple paths between states $z'$ and $z$ are possible, the RHS in~\eqref{eq:master_equation} needs to include a sum over all possible jump paths.}
	
	We assume that one can meaningfully divide the state space into distinct parts, e.g., $z = \{x,y\}$, where two subsystems $X$ and $Y$ are identified as distinct units interacting with each other. The process $z(t)$ is \emph{bipartite} if the transition rates can be written as
	\begin{equation}
		R(z|z';t) = R^{xx'}_{y}(t)\, \delta_{y,y'} + R^{x}_{yy'}(t)\, \delta_{x,x'}\,, \label{eq:bipartite_assumption}
	\end{equation}
	meaning that transitions cannot happen simultaneously in multiple subsystems.
	\blue{Note that this does not imply that the processes $x(t)$ and $y(t)$ are independent of each other; rather their influence on each other is restricted to modifying the other process's transition rates.}
	
	When the dynamics of the joint system are not bipartite, the dissection of energy and information flows presented in the following is more challenging. Ch\'etrite, \emph{et al}.\ have investigated this case~\cite{Chetrite2019_Information}.
	\blue{Moreover, information flows for quantum systems (without bipartite structure) have also been analyzed~\cite{Ptaszynski2019_Thermodynamics}.}
	\blue{Here, we exclusively cover classical bipartite systems.}

	\subsection{Paradigmatic examples}
	Bipartite dynamics should be expected whenever two systems (that possess their own dynamics) are combined such that each fluctuation can be decomposed into independent contributions. The dynamics of systems studied in cellular biology can often be approximated as bipartite.
	
\begin{figure*}[ht]
		\centering
		\includegraphics[width=0.4\linewidth]{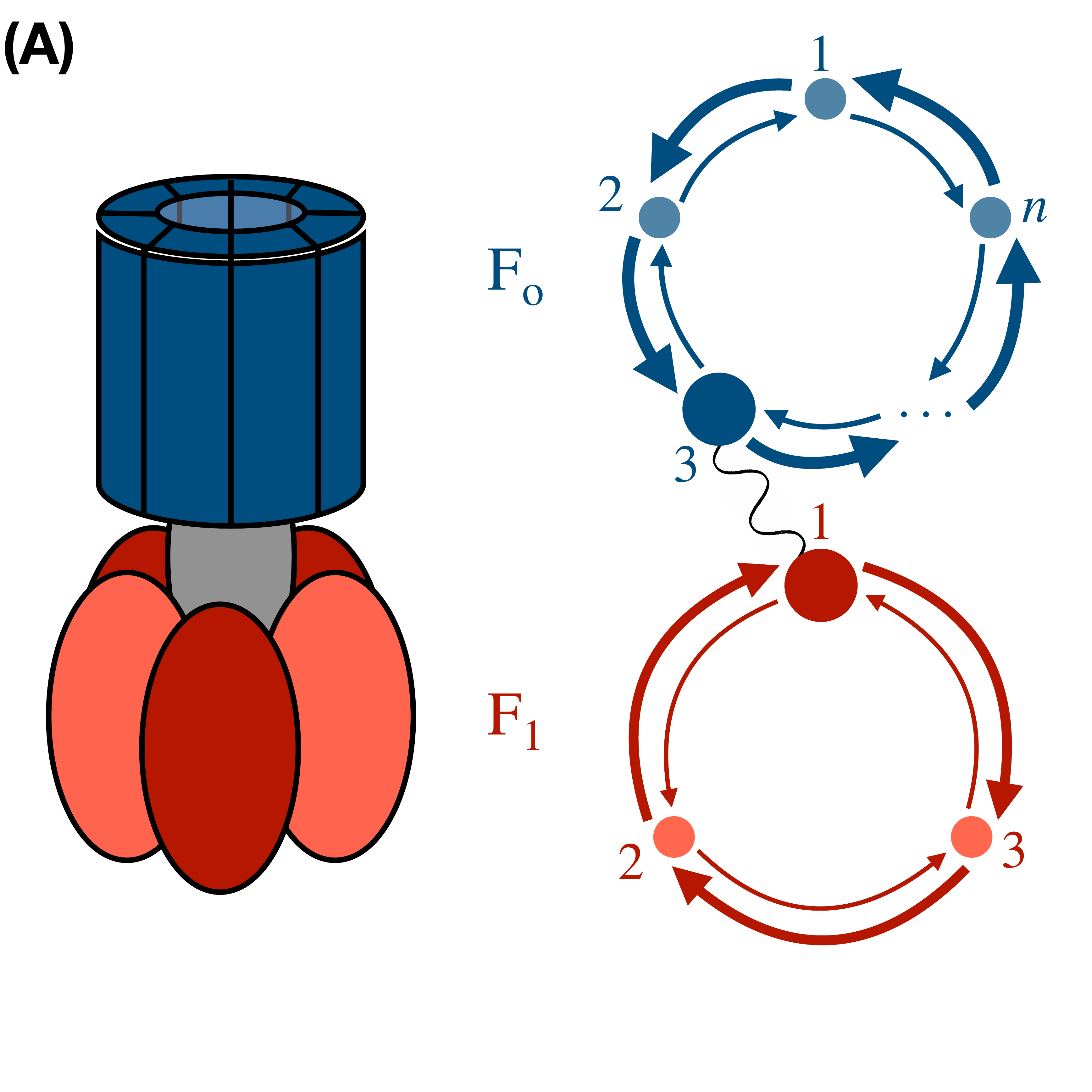}\hspace{1cm}
		\includegraphics[width=0.4\linewidth]{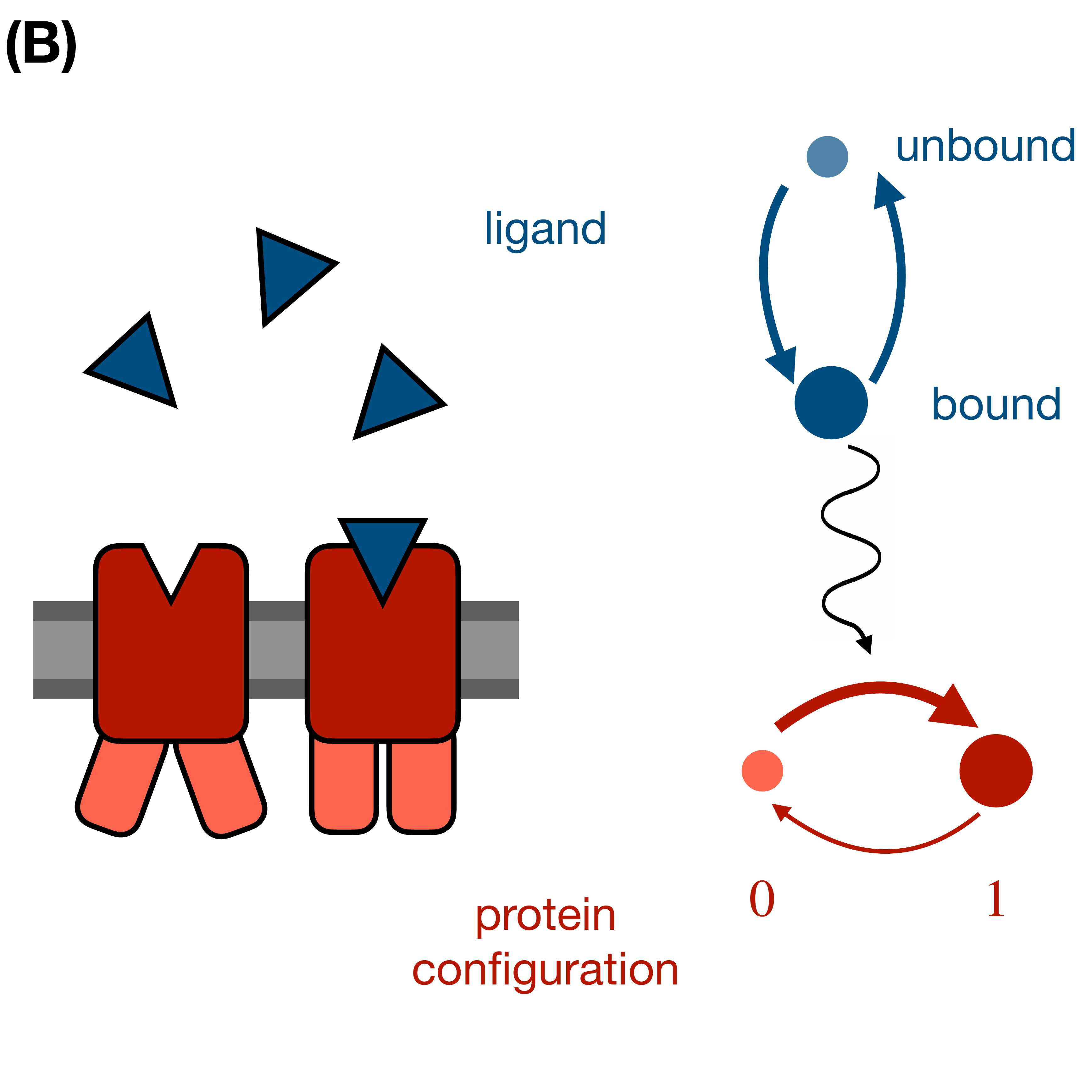}
		\caption{Paradigmatic example systems and their simplified state graphs. \textbf{(A)} Simplified model for $\mathrm{F}_{\rm o}\!-\!\mathrm{F}_1$ ATP synthase. Upstream $\mathrm{F}_{\rm o}$ dynamics are cyclically driven by a proton gradient while downstream $\mathrm{F}_1$ dynamics are driven in the opposite direction by ATP hydrolysis~\cite{Junge2015_ATP_Synthase,Lathouwers2020_Nonequilibrium,Lathouwers2022_Internal}. Through their coupling, the joint system can transduce work by driving the downstream system against its natural gradient, thereby converting one chemical fuel into another. \textbf{(B)} Simplified model of a biochemical sensor, e.g., involved in \emph{E.~coli} chemotaxis~\cite{Barkai1997_Robustness,Sourjik2012_Responding,Mehta2012_Energetic,Barato2013_Rate,Barato2013_Information-theoretic,Barato2014_Efficiency}. The upstream signal is the binding state (bound or unbound) of the receptor which is reflected in the downstream protein conformation by modifying its potential-energy landscape and thereby influencing the transition rates between configurations.}
		\label{fig:paradigmatic_examples}
	\end{figure*}

	Two paradigmatic examples that have been well studied are molecular motors (such as $\mathrm{F}_\mathrm{o}\!-\!\mathrm{F}_1$ ATP synthase) with strongly coupled interacting sub-components, or cellular sensors that react to a changing external concentration. The joint dynamics of such systems can be decomposed into the distinct fluctuations of each subsystem, each of which is influenced by the other subsystem (in the case of a strongly coupled molecular machine) or into dynamics strongly influenced by an independent process (in the case of a sensor). Figure~\ref{fig:paradigmatic_examples} shows examples and associated simplified state graphs.
	
	\subsection{Notation}
	To keep the notation concise and unambiguous, we adopt the following conventions:
	\begin{enumerate}
		\item \blue{Random variables are denoted with small letters. Occasionally the more explicit notation $p(X_t\!=\!x)$ is used to avoid ambiguity.}
		\item \blue{The joint probability of two random 
			variables
			taking values $x$ and $y$, respectively, is denoted by $p(x,y)$. The conditional probability of $x$ given $y$ is denoted by $p(x|y)$.}
		\item Time arguments are dropped for probabilities and transition rates unless distinct times appear in a single expression, as in $p(x_t,y_{t'})$.
		\item Total time derivatives are denoted with 
		\blue{a dot. The bipartite assumption ensures that rates of change of various quantities can be split into separate contributions due to the $X$ and $Y$ dynamics, respectively. Those individual rates of change}
		are indicated with a dot and the corresponding superscript, i.e., $\dot E^X$ 
		\blue{is the}
		rate of change of energy ($E$) due to $X$-dynamics and 
		\blue{$\dot E = \dot E^X + \dot E^Y$ [see~\eqref{eq:split_energy_balance}-\eqref{eq:energy_change_y}].}
		\item When no argument is given, symbols represent \emph{global quantities}, whereas capitalized arguments in square brackets indicate different  \emph{subsystem-specific} quantities, e.g., $S:=-\sum_{x,y} p(x,y)\ln p(x,y)$ is the joint entropy, while $S[X] := -\sum_{x} p(x)\ln p(x)$ and $S[X|Y] := -\sum_{xy} p(x,y)\ln p(x|y)$ are marginal and conditional entropies, respectively~\cite[Chap. 2]{Cover2006_Elements}.
	\end{enumerate}

	\subsection{Continuous state spaces} \label{sec:continuous_state-spaces}
	The framework outlined below can also be applied to continuous state spaces. For continuous diffusion processes described by a Fokker-Planck equation~\cite{Risken1996_Fokker-Planck} this has been done in \cite{Horowitz2015_Multipartite}.
	
	For diffusion-type dynamics, \eqref{eq:bipartite_assumption} corresponds to the statement that the diffusion matrix must be block-diagonal, such that the Fokker-Planck equation can be written as
	\begin{align}
		\frac{\partial p_t(x,y)}{\partial t} = &-\frac{\partial}{\partial x} \left[\mu^X(x,y;t)- D^X \frac{\partial}{\partial x}\right]\,p_t(x,y)\\
		&\qquad-\frac{\partial}{\partial y} \left[\mu^Y(x,y;t)-D^Y \frac{\partial}{\partial y}\right]\,p_t(x,y)\,, \nonumber
	\end{align}
	for respective subsystem drift coefficients $\mu^X$ and $\mu^Y$ and subsystem diffusion coefficients $D^X$ and $D^Y$. The corresponding coupled Langevin equations~\cite{Gardiner2004_Handbook} are
	\begin{subequations}
		\begin{align}
			\dot x &= \mu^X(x,y;t) + \sqrt{D^X}\,\xi^X(t) \label{eq:bipartite_Langevin_x}\\
			\dot y &= \mu^Y(x,y;t) + \sqrt{D^Y}\,\xi^Y(t) \label{eq:bipartite_Langevin_y}\,,
		\end{align}
	\end{subequations}%
	where $\xi^X(t)$ and $\xi^Y(t)$ are independent Gaussian white-noise terms for which $\left\langle \xi^X(t) \xi^X(t') \right\rangle \!=\! 2 \delta(t-t')$, and similarly for $Y$, and $\left\langle \xi^X(t) \xi^Y(t') \right\rangle \!=\! 0$. Therefore the two components $X$ and $Y$ are indeed influenced by independent fluctuations, which is often a reasonable approximation for most systems studied here, e.g., two-component molecular machines.

	\section{Energy flows} \label{sec:energy_flows}
	As a first step towards a thermodynamic interpretation of the stochastic dynamics described above, we relate stochastic transitions to energy exchanges between the two subsystems and between individual subsystems and the environment as represented by baths/reservoirs of various kinds. For the systems considered here it is safe to assume that all processes are \emph{isothermal} and that their stochasticity is due to thermal fluctuations.
	
	For systems relaxing to equilibrium the transition rates in \eqref{eq:master_equation} and \eqref{eq:bipartite_assumption} are related to thermodynamic potentials through the \emph{detailed-balance relation}. This relation follows from demanding that, in the absence of any driving, the distribution of system states must relax to the equilibrium distribution with no net flux along any transition,
	\begin{equation}
		0 = R^{xx'}_{yy'} p_\mathrm{eq}(x',y') - R^{x'x}_{y'y} p_\mathrm{eq}(x,y)\,. \label{eq:flux_vanish_eq}
	\end{equation}
	The equilibrium distribution is the Boltzmann distribution $p_\mathrm{eq}(x,y)=\exp\left[ -\left(\epsilon_{xy}-F_\mathrm{eq} \right)/k_\mathrm{B}T\right]$, where $\epsilon_{xy}$ is the potential energy of the system state $(x,y)$, $k_\mathrm{B}$ is Boltzmann's constant, $T$ the temperature, and $F_\mathrm{eq}$ the equilibrium free energy. Consequently, the transition rates are related by
	\begin{equation}
		\blue{\ln\frac{R^{xx'}_{yy'}}{R^{x'x}_{y'y}}} 
		= \frac{\epsilon_{x'y'}-\epsilon_{xy}}{k_\mathrm{B}T}\,.
	\end{equation}
	\blue{When each \emph{system state} is a mesostate composed of many microstates, as is common for modeling small biological systems, the potential energy $\epsilon_{xy}$ must be replaced by a mesostate \emph{free energy}~\cite{Seifert2019_From}. The following thermodynamic formalism remains unchanged, however.}
	
	The systems we consider here are driven by chemical reactions and external loads and do not obey the detailed-balance relation. Consequently they do not, in general, relax to equilibrium. Conceptually, we could include the state of the other reservoirs (chemical and work reservoirs) into the microstate $Z$ of the system and then describe a nonequilibrium steady state as a very slow relaxation to global equilibrium, driving cyclical processes in the system of interest; however, such a description would be unnecessarily cumbersome. Assuming that these reservoirs are large compared to the system of interest and weakly coupled to it, we split the free energy of the supersystem into contributions from the reservoirs and the system of interest. Then, energy exchanges between all reservoirs and the system of interest are treated in the same way as energy exchanges with a heat bath, giving
	\blue{a \emph{local}}
	detailed-balance relation~\cite{Bergmann1955_New,VandenBroeck2015_Ensemble,Seifert2019_From,Maes2021_Local}:
	\begin{align}
		\ln\frac{R^{xx'}_{yy'}}{R^{x'x}_{y'y}} &= \label{eq:generalized_detailed_balance}\\
		&\!\!\!\!\!\!\!\!\!\!\frac{\epsilon_{x'y'}-\epsilon_{xy} - \sum_\nu \left(\Delta\mu^{xx'}_{yy'}\right)^{(\nu)} + f^{X}\,\Delta^{xx'} + f^{Y}\,\Delta_{yy'}}{k_\mathrm{B}T}\,,\nonumber
	\end{align}
	where $\left(\Delta\mu^{xx'}_{yy'}\right)^{(\nu)}$ is the free-energy change in the reservoir $\nu$ associated with system transition $(x',y') \to (x,y)$, $f^{X}$ and $f^{Y}$ are external forces (here assumed constant) acting on the respective subsystems, and $\Delta^{xx'}$ and $\Delta_{yy'}$ are the respective lengths the subsystems undertake against their respective external forces when stepping from $(x',y')$ to $(x,y)$.
	\blue{The RHS of Eq.~\eqref{eq:generalized_detailed_balance} is sometimes called \emph{entropy flow} associated with a transition~\cite{VandenBroeck2015_Ensemble}.}
	\blue{Importantly, if multiple paths connect the states $\{x',y'\}$ and $\{x,y\}$, local detailed-balance relations hold separately for each of these paths~\cite{Esposito2012_Coarse_graining}.}

	\subsection{Global energy balance}
	Armed with the
	\blue{local}
	detailed-balance relation~\eqref{eq:generalized_detailed_balance}, we identify different energy flows in the system.
	\blue{Below, we state the usual conventions of}
	stochastic thermodynamics
	\blue{~\cite{Seifert2012_Stochastic,VandenBroeck2015_Ensemble} to}
	identify the different contributions (heat and work) associated with each transition. 
	The average rate $\dot Q(t)$ of \emph{heat} exchanged with the thermal environment is given by averaging the log-ratio of transition rates over the net flux for all transitions in the system:
	\begin{equation}
		\dot Q := - k_\mathrm{B} T \!\!\!\!\sum_{x>x',y>y'}\!\!\!\! \left[ R^{xx'}_{yy'}\, p(x',y') - R^{x'x}_{y'y}\, p(x,y)\right]\ln\frac{R^{xx'}_{yy'}}{R^{x'x}_{y'y}} \ , \label{eq:rate_heat}
	\end{equation}
	\blue{where we assume that the states $x$ and $y$ are consecutively numbered, so that the notation $x>x'$ indicates a sum over transitions between distinct states, omitting the reverse transitions.}
	Throughout this review all energy flows into the system are positive by convention.
	
	Two types of work can be identified, $\dot W = \dot W_\mathrm{chem} + \dot W_\mathrm{mech}$. These are the average rate of \emph{chemical work} associated with the influx of energy from the chemical reservoirs,
	\begin{align}
		\dot W_\mathrm{chem} &:= -\!\!\!\! \sum_{x>x',y>y'} \!\!\!\!\left[ R^{xx'}_{yy'}\, p(x',y') - R^{x'x}_{y'y}\, p(x,y)\right]\nonumber\\
		&\qquad\qquad\qquad\qquad\times \sum_\nu \left(\Delta\mu^{xx'}_{yy'}\right)^{(\nu)} \label{eq:rate_chemical_work} \,,
	\end{align} 
	and the average rate of \emph{mechanical work} due to the subsystems' responses to external forces,
	\begin{align}
		\dot W_\mathrm{mech} &:= \sum_{x>x',y>y'}\!\!\!\! \left[ R^{xx'}_{yy'}\, p(x',y') - R^{x'x}_{y'y}\, p(x,y)\right]\nonumber\\
		&\qquad\qquad\qquad\times\left(f^{X}\Delta^{xx'} + f^{Y}\Delta_{yy'}\right) \label{eq:rate_mechanical_work} \,.
	\end{align} 
	
	Finally, we identify the rate of change of \emph{average internal energy} as
	\begin{equation}
		\dot E := \!\!\sum_{x>x',y>y'}\!\! \left[ R^{xx'}_{yy'}\, p(x',y') - R^{x'x}_{y'y}\, p(x,y)\right]\left(\epsilon_{xy} - \epsilon_{x'y'} \right)\,. \label{eq:rate_internal_energy}
	\end{equation} 
	
	With \eqref{eq:generalized_detailed_balance}-\eqref{eq:rate_mechanical_work}, we verify the \emph{global first law}, representing the global energy balance:
	\begin{align}
		\dot E = \dot Q + \dot W_\mathrm{chem} + \dot W_\mathrm{mech}\,, \label{eq:global_first_law}
	\end{align} 
	which retrospectively justifies identifying the log-ratio of transition rates as heat~\eqref{eq:rate_heat}.

	\blue{\subsection{Subsystem-specific energy balances}}
	Due to the bipartite assumption~\eqref{eq:bipartite_assumption}, we also find 
	\blue{subsystem-specific}
	versions of this balance equation by splitting all energy flows into contributions from the respective subsystems: First the heat flow
	\begin{subequations}
		\begin{equation}
			\dot Q = \dot Q^X + \dot Q^Y\, \label{eq:split_heat}
		\end{equation}
		splits into subsystem-specific heat flows
		\begin{align}
			\dot Q^X &:= - k_\mathrm{B} T\! \sum_{x>x',y} \left[ R^{xx'}_{y}\, p(x',y) - R^{x'x}_{y}\, p(x,y)\right]\,\ln\frac{R^{xx'}_{y}}{R^{x'x}_{y}} \label{eq:rate_heat_x}\\
			\dot Q^Y &:= - k_\mathrm{B} T\! \sum_{x,y>y'} \left[ R^{x}_{yy'}\, p(x,y') - R^{x}_{y'y}\, p(x,y)\right]\ln\frac{R^{x}_{yy'}}{R^{x}_{y'y}}\,. \label{eq:rate_heat_y}
		\end{align}
	\end{subequations}
	Similarly, the chemical work
	\begin{subequations}
		\begin{equation}
			\dot W_\mathrm{chem} = \dot W_\mathrm{chem}^X + \dot W_\mathrm{chem}^Y
		\end{equation}
		splits into
		\begin{align}
			\dot W_\mathrm{chem}^X &:= - \sum_{x>x',y} \left[ R^{xx'}_{y}\, p(x',y) - R^{x'x}_{y}\, p(x,y)\right]\nonumber\\
			&\qquad\qquad\qquad\qquad\times\sum_\nu \left(\Delta\mu^{xx'}_y\right)^{(\nu)} \label{eq:chemical_power_x}\\
			\dot W_\mathrm{chem}^Y &:= - \sum_{x,y>y'} \left[ R^{x}_{yy'}\, p(x,y') - R^{x}_{y'y}\, p(x,y)\right]\nonumber\\
			&\qquad\qquad\qquad\qquad\times \sum_\nu \left(\Delta\mu^{x}_{yy'}\right)^{(\nu)}\,, \label{eq:chemical_power_y}
		\end{align}
	\end{subequations}
	\blue{where $\Delta\mu^{xx'}_y$ is equal to $\Delta\mu^{xx'}_{yy'}$ evaluated for $y=y'$ and similarly for $\Delta\mu^x_{yy'}$. The bipartite assumption~\eqref{eq:bipartite_assumption} ensures that these two functions together cover all applicable $\Delta\mu^{xx'}_{yy'}$.}
	
	Finally, the mechanical work
	\begin{subequations}
		\begin{equation}
			\dot W_\mathrm{mech} = \dot W_\mathrm{mech}^X + \dot W_\mathrm{mech}^Y
		\end{equation}
		splits into
		\begin{align}
			\dot W_\mathrm{mech}^X &:= \sum_{x>x',y} \left[ R^{xx'}_{y}\, p_{t}(x',y) - R^{x'x}_{y}\, p(x,y)\right]\, 
			\blue{f^X}
			\Delta^{xx'} \\
			\dot W_\mathrm{mech}^Y &:= \sum_{x,y>y'} \left[ R^{x}_{yy'}\, p_{t}(x,y') - R^{x}_{y'y}\, p_{t}(x,y)\right]\, 
			\blue{f^Y}
			\Delta_{yy'} \,.
		\end{align}
	\end{subequations}
	
	Moreover, we formally split the change in the joint potential energy 
	\begin{subequations}
		\begin{equation}
			\dot E = \dot E^X + \dot E^Y \label{eq:split_energy_balance}
		\end{equation}
		into contributions due to the respective dynamics of each particular subsystem,
		\begin{align}
			\dot E^X &:= \sum_{x>x',y} \left[ R^{xx'}_{y}\, p(x',y) - R^{x'x}_{y}\, p(x,y)\right]\,\left(\epsilon_{xy} - \epsilon_{x'y} \right) \label{eq:energy_change_x}\\
			\dot E^Y &:= \sum_{x,y>y'} \left[ R^{x}_{yy'}\, p(x,y') - R^{x}_{y'y}\, p(x,y)\right]\,\left(\epsilon_{xy} - \epsilon_{xy'} \right)\,, \label{eq:energy_change_y}
		\end{align}
	\end{subequations}
	\blue{where a positive rate indicates that the joint potential energy increases due to the respective subsystem's dynamics.}
	
	We obtain the 
	\blue{subsystem-specific}
	first laws as the 
	\blue{balances of energy flows into the respective subsystems:}
	\begin{subequations}
		\begin{align}
			\dot E^X &= \dot Q^X + \dot W_\mathrm{chem}^X + \dot W_\mathrm{mech}^X \label{eq:local_first_law_x} \\
			\dot E^Y &= \dot Q^Y + \dot W_\mathrm{chem}^Y + \dot W_\mathrm{mech}^Y \label{eq:local_first_law_y}\,.
		\end{align}
	\end{subequations}
	
	With \eqref{eq:bipartite_assumption}, \eqref{eq:generalized_detailed_balance}-\eqref{eq:rate_internal_energy}, and \eqref{eq:split_heat}-\eqref{eq:energy_change_y}, we verify that the sum of the
	\blue{subsystem-specific}
	first laws \eqref{eq:local_first_law_x} and \eqref{eq:local_first_law_y} yields the global first law~\eqref{eq:global_first_law}.

	\subsection{Work done by one subsystem on the other}\label{sec:transduced_work}
	The 
	\blue{subsystem-specific}
	first laws in~\eqref{eq:local_first_law_x} and~\eqref{eq:local_first_law_y} stem from a formal argument. Ideally, we would like to identify internal energy flows that the subsystems communicate between each other; i.e., we would like to identify \emph{transduced work} in the manner of~\cite{Large2021_Free-energy}. However, with no clear prescription on how to split the energy landscape into $X$-, $Y$-, and interaction components,
	\begin{equation}
		\epsilon_{xy} = \epsilon_x + \epsilon_y + \epsilon^\mathrm{int}_{xy}\,,
	\end{equation}
	the identification of energy flowing from one subsystem to the other remains ambiguous: For example, how much has a change in the $X$-coordinate changed the potential energy of the $X$-subsystem and how much has it changed the interaction energy? The ambiguity has already been pointed out in~\cite{Allahverdyan2009_Thermodynamic}, where the authors propose to settle it through physical arguments by identifying a clear interaction term in the Hamiltonian and asking that the splitting leaves constant the average subsystem energy.
	
	We propose a different approach to define an input work into one subsystem. Conventionally work is defined for interactions between a work reservoir (e.g., an experimentalist's external power source) and a system. Interactions between the work reservoir and the system are mediated by a \emph{control parameter} influencing the system's potential-energy landscape. Crucially, there is negligible feedback from the system state to the dynamics of the control parameter. To define work between subsystems, imagine treating subsystem $Y$ as if it were an externally manipulated control parameter influencing the potential-energy landscape of $X$. Then, the power done by the control parameter $Y$ on the system $X$ would be supplied externally and equal the rate of change of internal energy due to the dynamics of the control parameter:
	\begin{subequations}
		\begin{equation}
			\dot W^{Y \to X} 
			\blue{:=}
			\dot E^Y\,. \label{eq:trans_work_y_to_x}
		\end{equation}
		Consequently, we define $\dot W^{Y \to X}$ as the \emph{transduced work} from $Y$ to $X$,
		\blue{which is positive when $Y$ increases the potential energy available to $X$.}
		\blue{Similarly we define}
		\begin{equation}
			\dot W^{X \to Y} 
			\blue{:=} \dot E^X \label{eq:trans_work_x_to_y}
		\end{equation}
	\end{subequations}
	as the transduced work from $X$ to $Y$. Thus, an externally manipulated control parameter could be understood as the limiting case of a negligible back-action from the downstream system to the (possibly deterministic) dynamics of the upstream system. This identification of energy flows communicated between the systems becomes useful when singling out one subsystem that is driven (possibly with feedback) by another one (see section~\ref{sec:engine_setups}).
	
\begin{figure*}[ht]
		\centering
		\includegraphics[width=0.4\linewidth]{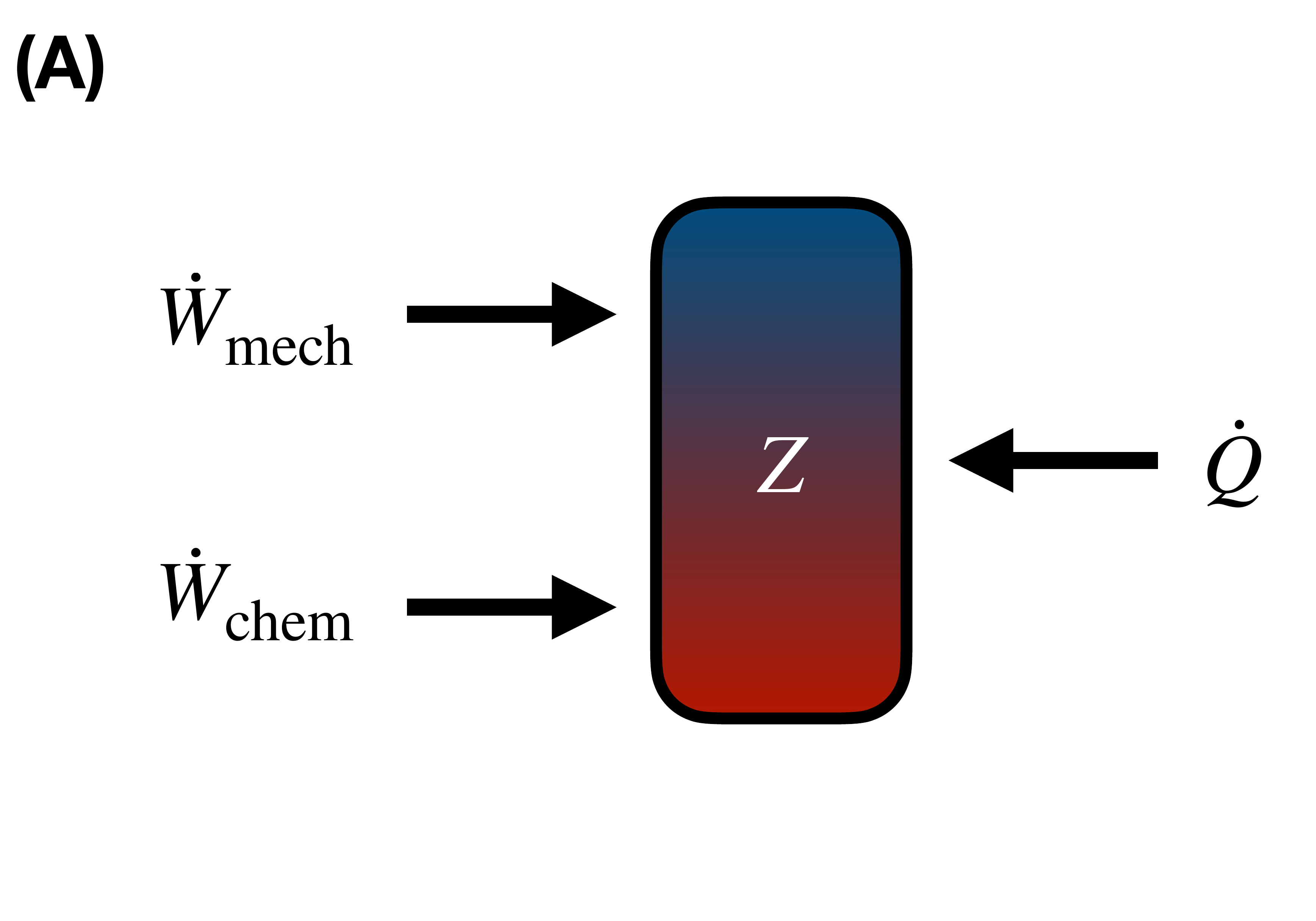}\hspace{1cm}
		\includegraphics[width=0.4\linewidth]{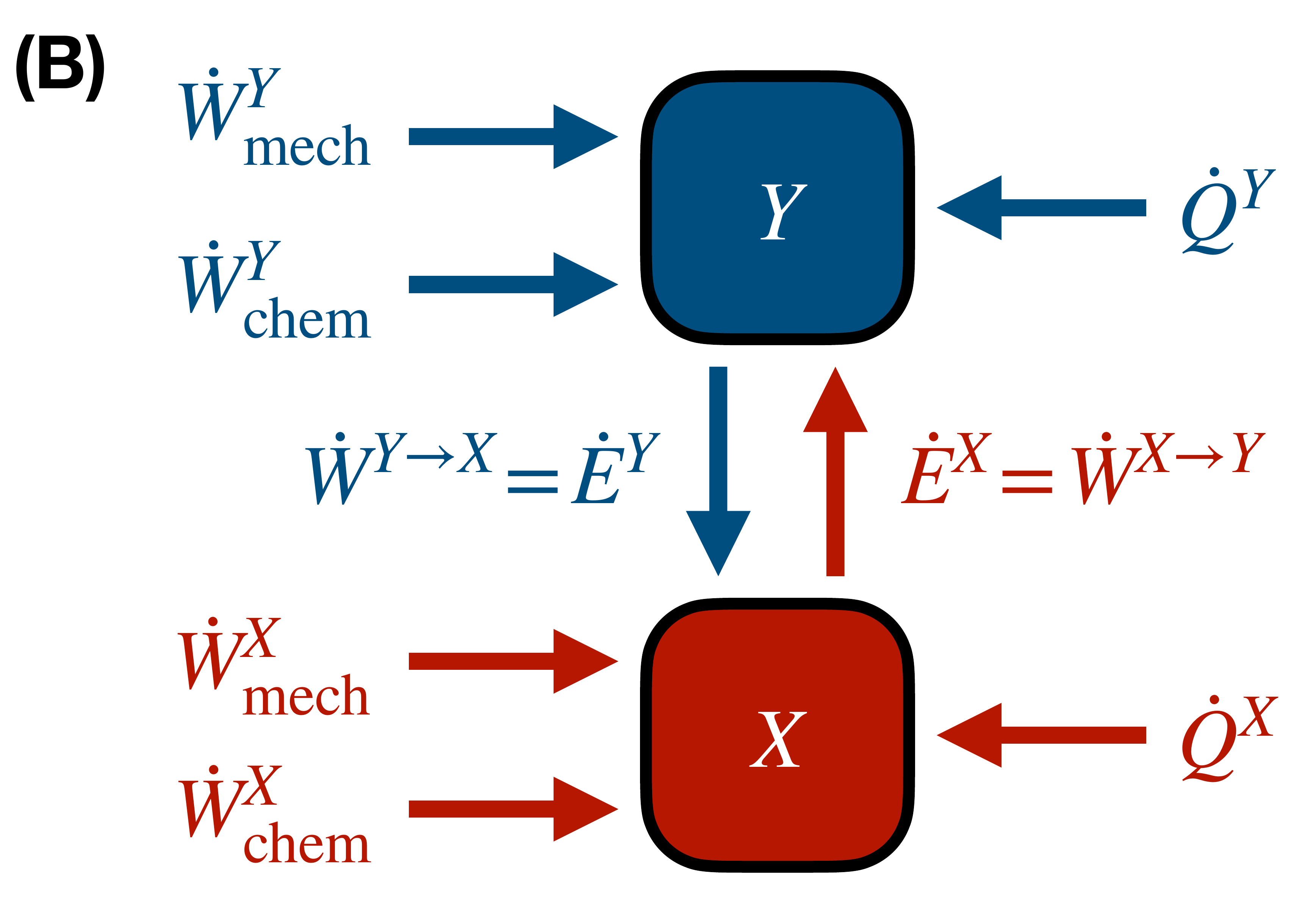}
		\caption{Energy flows in autonomous bipartite systems. \textbf{(A)} Global energy flows can be distinguished between work (mechanical and chemical) and heat. Since at steady state the average global internal energy stays constant, average flows of work and heat must cancel. \textbf{(B)} The bipartite assumption~\eqref{eq:bipartite_assumption} allows decomposition of energy flows into contributions from each subsystem. Color and direction of arrows reflect the
			\blue{subsystem-specific}
			first laws \eqref{eq:local_first_law_x} in red and \eqref{eq:local_first_law_y} in blue.}
		\label{fig:energy_flows}
	\end{figure*}

	Figure~\ref{fig:energy_flows} summarizes the splitting of the first law presented in this section and illustrates how energy moves between the subsystems.
	
	\section{Entropy and free-energy balance}\label{sec:entropy_balance}
	As always in thermodynamics, energetics are only half of the picture. We therefore next consider entropic quantities. Together, the rate
	\blue{$\dot E$}
	of change of global internal energy $E$~\eqref{eq:rate_internal_energy}, and the rate 
	\blue{$\dot S$}
	of change of global entropy $S$ (defined in the following) determine the rate of change of nonequilibrium (or ``generalized'') free energy~\cite{Gaveau1997_general_framework, Gaveau2008_Work_and_power, Esposito2011_Nonequilibrium_Free_Energy,Sivak2012_Near_equilibrium}, 
	\begin{equation} 
		\blue{\dot F := \dot E - k_\mathrm{B}T \dot S\,.}
		\label{eq:rate_global_free_energy}
	\end{equation}

	\subsection{Global entropy balance}
	Following Ref.~\cite{VandenBroeck2015_Ensemble}, we explicitly write the rate of change of system entropy:
	\begin{subequations}
		\begin{align}
			\blue{\dot S} &= -\sum_{x,y} \frac{\mathrm{d}}{\mathrm{d}t} p(x,y) \ln p(x,y)\\
			&= -\sum_{x, x', y,y'} \left[ R^{xx'}_{yy'}\, p(x',y') - R^{x'x}_{y'y}\, p(x,y) \right] \ln p(x,y) \label{eq:global_EP_balance_2}\\
			&= -\sum_{x>x',y>y'} \left[ R^{xx'}_{yy'} p(x',y') - R^{x'x}_{y'y} p(x,y) \right] \ln\frac{p(x,y)}{p(x',y')} \label{eq:global_EP_balance_3}\\
			&=\!\! \underbrace{\sum_{x>x',y>y'}\!\!\! \left[ R^{xx'}_{yy'} p(x',y') - R^{x'x}_{y'y} p(x,y) \right] \ln\frac{R^{xx'}_{yy'}\,p(x',y')}{R^{x'x}_{y'y}\,p(x,y)}}_{=: \dot\Sigma} \nonumber\\
			&\qquad- \underbrace{ \sum_{x>x',y>y'} \left[ R^{xx'}_{yy'} p(x',y') - R^{x'x}_{y'y} p(x,y) \right] \ln\frac{R^{xx'}_{yy'}}{R^{x'x}_{y'y}}}_{=-\dot Q/k_\mathrm{B}T}\,, \label{eq:global_EP_balance_4}
		\end{align}
	\end{subequations}
	where we have used the Master equation~\eqref{eq:master_equation} in~\eqref{eq:global_EP_balance_2},
	\blue{the fact that~\eqref{eq:global_EP_balance_2} sums over every transition twice in~\eqref{eq:global_EP_balance_3},}
	and the definition of the heat flow~\eqref{eq:rate_heat} in \eqref{eq:global_EP_balance_4}.
	
	Rearranging the terms gives the \emph{global second law}:
	\begin{subequations}
		\begin{align}
			\dot{\Sigma} &= 
			\blue{\dot S}
			- \frac{\dot Q}{k_\mathrm{B}T}  \label{eq:global_second_law_1}\\
			&=\!\!\!\sum_{x>x',y>y'}\!\!\! \left[ R^{xx'}_{yy'} p(x',y') - R^{x'x}_{y'y} p(x,y) \right] \ln\frac{R^{xx'}_{yy'}\,p(x',y')}{R^{x'x}_{y'y}\,p(x,y)} \label{eq:global_second_law_2}\\
			&\geq 0\,, \label{eq:global_second_law_3}
		\end{align}
	\end{subequations}
	where $\dot\Sigma$ is the rate of global entropy production, i.e., the rate at which entropy is produced in the whole system and attached baths. Its non-negativity follows from the fact that, in each term of the sum in \eqref{eq:global_second_law_2}, the two factors are always either both positive or both negative. 
	
	Using the definition of nonequilibrium free energy~\eqref{eq:rate_global_free_energy} and the global first law~\eqref{eq:global_first_law}, we rewrite the global second law as
	\begin{align}
		\dot W_\mathrm{mech} + \dot{W}_\mathrm{chem} - 
		\blue{\dot F \geq 0\,.}
	\end{align}

	\subsubsection{Marginal and hidden entropy production}
	An interesting digression
	\blue{covers}
	related research on inferring total entropy production from the dynamics of only one subsystem. Prominent examples of such systems with \emph{hidden degrees of freedom} are molecular transport-motor experiments~\cite{Kolomeisky2013_Motor,Zimmermann2015_Effective} in which only trajectories of an attached cargo are observed while the motor dynamics are hidden. Assessing motor efficiency, however, necessitates a detailed knowledge of the internal motor dynamics. Hence \emph{thermodynamic inference}~\cite{Seifert2019_From} is required to infer hidden system properties.
	
	Alongside dynamics on masked Markovian networks~\cite{Shiraishi2015_Fluctuation,Polettini2017_Effective,Bisker2017_Hierarchical,Martinez2019_Inferring,Skinner2021_Improved,Ehrich2021_Tightest,Skinner2021_Estimating,Hartich2022_Violation,vanderMeer2022_Thermodynamic,Harunari2022_What}, bipartite systems have been used to model situations in which one sub-component of the full system is hidden. One common strategy consists of mapping the observed dynamics of one subsystem onto a Markov model, which generally produces a lower bound on the total entropy production rate~\cite{Esposito2012_Coarse_graining} that can subsequently be augmented with any information available about the hidden dynamics~\cite{Large2021_Hidden}. However, the observed process is non-Markovian which results, e.g., in modifications of fluctuation theorems~\cite{Mehl2012_Role,Uhl2018_Fluctuations,Kahlen2018_Hidden}. Another approach is to use \emph{thermodynamic uncertainty relations}~\cite{Barato2015_Thermodynamic_Uncertainty, Gingrich2016_Dissipation_bounds,Li2019_Quantifying_dissipation,Manikandan2020_Short_Experiments,Vu2020_Optimal_current} to produce a bound on the total entropy production using observable currents.
	
	Importantly, the formalism laid out here requires full observability of the dynamics of both subsystems; recent efforts have explored when one can infer the kind of driving mechanism from observations of just one degree of freedom, e.g., the dynamics of probe particles attached to unobserved molecular motors~\cite{Pietzonka2014_Fine-structured,Zimmermann2015_Effective}.

	\blue{\subsection{Subsystem-specific entropy balance}}
	In analogy to
	\blue{subsystem-specific}
	versions of the first law~(\ref{eq:local_first_law_x},\ref{eq:local_first_law_y}) which introduce energy flows between the different subsystems,
	\blue{subsystem-specific}
	versions of the second law introduce an entropic flow between the systems, called \emph{information flow}.
	
	The bipartite assumption~\eqref{eq:bipartite_assumption} splits the global entropy production~\eqref{eq:global_second_law_2} into two nonnegative contributions, $\dot\Sigma = \dot\Sigma^X  + \dot\Sigma^Y$: 
	\begin{subequations}
		\begin{align}
			\blue{\dot\Sigma^X} &:=  \sum_{x>x',y} \left[ R^{xx'}_y p(x',y) - R^{x'x}_y p(x,y) \right] \ln\frac{R^{xx'}_y p(x',y)}{R^{x'x}_y p(x,y)}\nonumber\\
			&\geq 0\\
			\blue{\dot\Sigma^Y} &:= \sum_{x,y>y'} \left[ R^{x}_{yy'} p(x,y') - R^{x}_{y'y} p(x,y) \right] \ln\frac{R^{x}_{yy'} p(x,y')}{R^{x}_{y'y} p(x,y)}\nonumber\\
			&\geq 0\,.
		\end{align}
	\end{subequations}
	
	To make contact with the form of the global second law~\eqref{eq:global_second_law_1}, we identify different contributions to the \emph{
		\blue{subsystem-specific}
		entropy productions} $\dot\Sigma^X$ and $\dot\Sigma^Y$:
	\begin{subequations}
		\begin{align}
			\dot\Sigma^X &= \dot S^X - \frac{\dot Q^X}{k_\mathrm{B}T} \label{eq:local_EP_x}\\
			\dot\Sigma^Y &= \dot S^Y - \frac{\dot Q^Y}{k_\mathrm{B}T}\,, \label{eq:local_EP_y}
		\end{align}
	\end{subequations}
	where $\dot Q^X$ and $\dot Q^Y$ are the
	heat flows~\eqref{eq:rate_heat_x}~and~\eqref{eq:rate_heat_y}
	\blue{into the respective subsystems}
	and, in accordance with the identification of rates of change of energy in~\eqref{eq:energy_change_x}~and~\eqref{eq:energy_change_y}, we identify the rates of change of \emph{total entropy} due to the \emph{individual subsystem dynamics}, 
	\begin{subequations} 
		\begin{align}
			\dot S^X &
			\blue{:=}
			- \sum_{x>x',y} \left[ R^{xx'}_y p(x',y) - R^{x'x}_y p(x,y) \right] \ln\frac{p(x,y)}{p(x',y)} \label{eq:rate_total_entropy_X}\\
			\dot S^Y &
			\blue{:=} - \sum_{x,y>y'} \left[ R^{x}_{yy'}\, p(x,y') - R^{x}_{y'y}\, p(x,y) \right] \ln\frac{p(x,y)}{p(x,y')}\,.
		\end{align}
	\end{subequations}
	Importantly, these rates \emph{are not} the rates of change of marginal entropies $S[X]=-\sum_x p(x)\ln p(x)$ and $S[Y]=-\sum_y p(y)\ln p(y)$. Rewriting the
	\blue{subsystem-specific}
	entropy productions with these marginal rates leads to the identification of an \emph{information flow}, as we show in section~\ref{sec:local_2nd_law_info_flow}.
	
	Substituting the
	\blue{subsystem-specific}
	first laws~(\ref{eq:local_first_law_x},\ref{eq:local_first_law_y}) gives
	\blue{subsystem-specific}
	second laws in terms of work and free energy,
	\begin{subequations}
		\begin{align}
			\dot W^X_\mathrm{mech} + \dot W^X_\mathrm{chem} - \dot F^X &\geq 0\\
			\dot W^Y_\mathrm{mech} + \dot W^Y_\mathrm{chem} - \dot F^Y &\geq 0\,,
		\end{align}
	\end{subequations}
	where $\dot F^X = \dot E^X - k_\mathrm{B}T \dot S^X$ is the rate of change of nonequilibrium free energy due to the dynamics of $X$, and similarly for $\dot F^Y$. Their sum gives the rate of change 
	\blue{$\dot F$}
	of the nonequilibrium free energy. 
	
	While formally appealing, the rate of change of nonequilibrium free energy due to one subsystem's dynamics has little utility. Often, one only knows the free energy for one subsystem (e.g., by having constructed a potential-energy landscape for one of the subsystems as done in~\cite{Kawaguchi2014_Nonequilibrium} for the $F_1$-component of ATP synthase) or the free energy of one subsystem is unknown or undefined (e.g., for the external environment process in a sensing setup). To this end, we next present other ways of writing (and interpreting) the
	\blue{subsystem-specific}
	second laws.

	\blue{\subsection{Subsystem-specific second laws with information flows}}
	\label{sec:local_2nd_law_info_flow}
	We express the rate $\dot S^X$ of change of total entropy due to the dynamics of $X$~\eqref{eq:rate_total_entropy_X} in terms of the rate of change of marginal entropy,
	\begin{align} 
		\blue{\dot S[X]}
		&= - \sum_{x>x',y} \left[ R^{xx'}_y p(x',y) - R^{x'x}_y p(x,y) \right] \ln\frac{p(x)}{p(x')}\,,
	\end{align}
	and similarly for $\dot S^Y$ to rewrite~(\ref{eq:local_EP_x},\ref{eq:local_EP_y}) as
	\begin{subequations}
		\begin{align}
			0 \leq \dot\Sigma^X &= 
			\blue{\dot S[X]}
			- \frac{\dot Q^X}{k_\mathrm{B}T} - \dot I^X \label{eq:local_EP_x_info_flow}\\
			0 \leq \dot\Sigma^Y &= 
			\blue{\dot S[Y]}
			- \frac{\dot Q^Y}{k_\mathrm{B}T} - \dot I^Y\,. \label{eq:local_EP_y_info_flow}
		\end{align}
	\end{subequations}
	Here, we have identified the remaining terms as the \emph{information flows}~\cite{Allahverdyan2009_Thermodynamic},
	\begin{subequations}
		\begin{align}
			\dot I^X &
			\blue{:=}
			\lim\limits_{\mathrm{d}t \to 0}\frac{I[X_{t+\mathrm{d}t};Y_t] - I[X_{t};Y_t]}{\mathrm{d}t} \label{eq:info_flow_change_mut_info_x}\\
			\dot I^Y &
			\blue{:=}
			\lim\limits_{\mathrm{d}t \to 0}\frac{I[X_{t};Y_{t+\mathrm{d}t}] - I[X_{t};Y_{t}]}{\mathrm{d}t}\,, \label{eq:info_flow_change_mut_info_y}
		\end{align}
	\end{subequations}
	i.e., the rate of change of mutual information between the subsystems that is due only to the dynamics of one of them. 
	\blue{Information flow is positive when the dynamics of the corresponding subsystem increase the mutual information between the two subsystems.}
	In Appendix~\ref{app:info_flows_bipartite_Markov} we show that for bipartite Markovian dynamics this definition leads to
	\begin{subequations}
		\begin{align}
			\dot I^X = \sum_{x>x',y} \left[ R^{xx'}_y p(x',y) - R^{x'x}_y p(x,y) \right] \ln\frac{p(y|x)}{p(y|x')}\,, \label{eq:info_flow_x_Markov}\\
			\dot I^Y = \sum_{x,y>y'} \left[ R^{x}_{yy'}\, p(x,y') - R^{x}_{y'y}\, p(x,y) \right] \ln\frac{p(x|y)}{p(x|y')}\,, \label{eq:info_flow_y_Markov}
		\end{align}
	\end{subequations}
	i.e., the form used in~\cite{Horowitz2014_Thermodynamics} with which we can verify the equality of~\eqref{eq:local_EP_x}~and~\eqref{eq:local_EP_x_info_flow} and similarly of~\eqref{eq:local_EP_y}~and~\eqref{eq:local_EP_y_info_flow}.
	\blue{Equations~(\ref{eq:local_EP_x_info_flow},\ref{eq:local_EP_y_info_flow}) express the same subsystem-specific entropy productions as Eqs.~(\ref{eq:local_EP_x},\ref{eq:local_EP_y}). The latter contain subsystem-specific changes $\dot S^X$ and $\dot S^Y$ of the global (joint) entropy $S$. In contrast, the former contain changes of the marginal entropies $S[X]$ and $S[Y]$. Joint entropy not only contains the sum of marginal entropies, but also the mutual information, $S = S[X]+ S[Y] - I[X;Y]$~\cite[Chap. 2.3]{Cover2006_Elements}. Consequently, changes in joint entropy not only contain contributions from the changes of marginal entropies, but also the change in mutual information. The information flows $I^X$ and $I^Y$ distribute this rate of change symmetrically across the two subsystem-specific entropy productions.}
	\blue{Summing~\eqref{eq:info_flow_change_mut_info_x}~and~\eqref{eq:info_flow_change_mut_info_y}}
	yields the total change in mutual information between $X$ and $Y$. For Markovian bipartite dynamics this reads explicitly
	\begin{widetext}
		\begin{subequations}
			\begin{align}
				\dot I^X + \dot I ^Y &= \sum_{x>x',y>y'} \left[R^{xx'}_{yy'} p(x',y') - R^{x'x}_{y'y} p(x,y) \right]
				\blue{\left[\ln\frac{p(y|x)}{p(y|x')} + \ln\frac{p(x|y)}{p(x|y')} \right]}\\
				&= \sum_{x>x',y>y'} \left[R^{xx'}_{yy'} p(x',y') - R^{x'x}_{y'y} p(x,y) \right]
				\blue{\Bigg[\ln\frac{p(x,y)}{p(x)p(y)}\frac{p(x')p(y')}{p(x',y')} + \ln\frac{p(x,y)}{p(x)p(y)}\frac{p(x')p(y')}{p(x',y')} \Bigg]}\\
				&=\sum_{x>x',y>y'} \left[R^{xx'}_{yy'} p(x',y') - R^{x'x}_{y'y} p(x,y) \right]\ln\frac{p(x,y)}{p(x)p(y)}\frac{p(x')p(y')}{p(x',y')}\\
				&= 
				\blue{\dot I\,,} \label{eq:sum_info_flows}
			\end{align}
		\end{subequations}%
	\end{widetext}
	\blue{where the bipartite assumption~\eqref{eq:bipartite_assumption} ensures that there is no contribution from transitions in which $x$ and $y$ change simultaneously.}
	
	The term \emph{information flow} was first used in the context of nonequilibrium thermodynamics by Allahverdyan \emph{et al.}~\cite{Allahverdyan2009_Thermodynamic} and was later taken up by Horowitz and Esposito~\cite{Horowitz2014_Thermodynamics}. 
	\blue{Section}
	~\ref{sec:Nostalgia} compares information flow with conceptually similar quantities called \emph{nostalgia}~\cite{Still2012_Prediction} and \emph{learning rate}~\cite{Barato2014_Efficiency}.
	
	Notice the appealing structure of the
	\blue{subsystem-specific}
	entropy productions in \eqref{eq:local_EP_x_info_flow} and \eqref{eq:local_EP_y_info_flow}: For interacting 
	\blue{subsystems},
	it is not enough to consider marginal entropy changes and heat flows into one 
	\blue{subsystem,}
	because to obtain a nonnegative entropy production rate, one needs an additional term due to correlations between the interacting 
	\blue{subsystems}.
	Expressed differently: When one explicitly neglects or is unaware of other subsystems strongly coupled to the 
	\blue{subsystem}
	of interest, erroneous conclusions about the entropy production are possible, either overestimating it or perhaps even finding it to be negative, leading to a Maxwell-demon-like paradox.
	
	We next present two alternative representations of the same
	\blue{subsystem-specific}
	entropy production that rely on rewriting the rate of change of global entropy $\dot S^X$ due to $X$ dynamics in terms of the time-derivative of conditional entropy,
	\blue{$\dot S[X|Y]$,}
	instead of the time-derivative of marginal entropy, 
	\blue{$\dot S[X]$.}

	\blue{\subsubsection{Alternative representation of subsystem-specific entropy production in terms of conditional entropy}}
	In addition to the formulation in~\eqref{eq:local_EP_x_info_flow}, the
	\blue{subsystem-specific}
	entropy production in~\eqref{eq:local_EP_x} can also be rewritten in terms of the rate of change 
	\blue{$\dot S[X|Y] = -\frac{\mathrm d}{\mathrm d t} \sum_{x,y} p(x,y) \ln p(x|y)$}
	of \emph{conditional entropy}, since
	\begin{subequations}
		\begin{align}
			\dot S^X &= - \sum_{x>x',y} \left[ R^{xx'}_y\, p(x',y) - R^{x'x}_y \,p(x,y) \right] \ln\frac{p(x,y)}{p(x',y)}\\
			&= -\!\! \sum_{x>x',y>y'}\!\! \left[ R^{xx'}_{yy'}\, p(x',y') - R^{x'x}_{y'y}\, p(x,y) \right] \ln\frac{p(x,y)}{p(x',y)} \label{eq:cond_entropy_info_flow_rel_2}\\
			&= -\!\!\! \sum_{x>x',y>y'}\!\!\! \left[ R^{xx'}_{yy'}\, p(x',y') - R^{x'x}_{y'y}\, p(x,y) \right]\nonumber\\
			&\qquad\qquad\times\ln\frac{p(x|y)}{p(x'|y')} \frac{p(x'|y')}{p(x'|y)}\\
			&= 
			\blue{\dot S[X|Y]}
			- \sum_{x',y>y'} \left[ R^{x'}_{yy'}\, p(x',y') - R^{x'}_{y'y}\, p(x',y) \right]\nonumber\\
			&\qquad\qquad\qquad\qquad\qquad\times\ln \frac{p(x'|y')}{p(x'|y)}\\
			&= 
			\blue{\dot S[X|Y]}
			+ \dot I^Y\,, \label{eq:cond_entropy_info_flow_rel}
		\end{align}
	\end{subequations}
	where
	\blue{line~\eqref{eq:cond_entropy_info_flow_rel_2} uses the bipartite assumption~\eqref{eq:bipartite_assumption} along with the fact that, due to the log-ratio in~\eqref{eq:cond_entropy_info_flow_rel_2}, all terms $y \neq y'$ are zero. Line~\eqref{eq:cond_entropy_info_flow_rel}}
	follows from the definition~\eqref{eq:info_flow_y_Markov} of $\dot I^Y$. This leads to
	\begin{align}
		0 \leq \dot\Sigma^X &= 
		\blue{\dot S[X|Y]}
		- \frac{\dot Q^X}{k_\mathrm{B}T} + \dot I^Y
		\label{eq:local_EP_x_cond_entropy}\,.
	\end{align}
	
	Comparing with~\eqref{eq:local_EP_x}, which expresses the same
	\blue{subsystem-specific}
	entropy production, we observe a difference in interpretation: If one interprets $Y$ not as a subsystem on equal footing with $X$ but instead as a \emph{stochastic control protocol} for the system $X$, the
	\blue{subsystem-specific}
	second law in \eqref{eq:local_EP_x_cond_entropy} seems more natural. Such stochastic control protocols arise naturally in the context of sensors, where a changing environment effectively acts as a stochastic protocol~\cite{Still2012_Prediction}, and in contexts with measurement-feedback loops where a stochastic measurement of the system state dictates the statistics of the future control protocol~\cite{Cao2009_Thermodynamics,Sagawa2010_Generalized,Ponmurugan2010_Generalized,Horowitz2010_Nonequilirbium,Sagawa2012_Nonequilibrium,Lahiri_2012}.

	\subsubsection{Subsystem-specific second law with conditional free energy}
	In cases where the nonequilibrium free energy of subsystem $X$ is known, we define a \emph{conditional nonequilibrium free energy} of system $X$ given a control parameter $Y$ as the average energy of $X$ given the particular control-parameter value $y$ less ($k_\mathrm B T$ times) the average entropy of $X$ given the control-parameter value $y$, all averaged over $Y$,
	\begin{subequations}
		\begin{align}
			F[X|Y] &:= \left\langle\!\! \left\langle \epsilon_{xy} \right\rangle_{p(x|y)}\! - k_\mathrm{B}T \left(\! - \sum_{x} p(x|y)\ln p(x|y)\! \right) \!\!\right\rangle_{p(y)}\\
			&= E - k_\mathrm{B} T S[X|Y]\,.
		\end{align}
	\end{subequations}
	Thus, this free energy is averaged over all stochastic control-parameter values.
	
	With the splitting of the first law in \eqref{eq:split_energy_balance}, the
	\blue{subsystem-specific}
	first law in  \eqref{eq:local_first_law_x}, and the identification of \emph{transduced work} $\dot W^{Y \to X}$ in \eqref{eq:trans_work_y_to_x}, we rewrite the
	\blue{subsystem-specific}
	second law in \eqref{eq:local_EP_x_cond_entropy} as
	\begin{subequations}
		\begin{align}
			0 \leq k_\mathrm{B} T\,\dot\Sigma^X &= \dot W^{Y \to X} + \dot W^X_\mathrm{mech} + \dot W^X_\mathrm{chem}\nonumber\\
			&\qquad\qquad\qquad -\blue{\dot F[X|Y]}
			+ k_\mathrm{B} T\, \dot I^Y \\
			&=
			\dot W^{\to X} - \dot F[X|Y]
			+ k_\mathrm{B} T\, \dot I^Y\,, \label{eq:2nd_law_cond_free_energy}
		\end{align}
	\end{subequations}
	where 
	\blue{$\dot W^{\to X} := \dot W^{Y \to X} + \dot W^X_\mathrm{mech} + \dot W^X_\mathrm{chem}$}
	is the total input work into subsystem $X$. Again, compared to the regular second law, there is an additional information flow modifying the entropy balance.

	\subsection{Steady-state flows}
	At steady state, average energy, entropy, and mutual information are all constant,
	\blue{$\dot E = \dot S = \dot S[X] = \dot S[X|Y] = \dot I[X;Y] =0$.}
	\blue{However, this does not imply that the subsystem-specific rates of change vanish, too; but the}
	energy and information flows do simplify, giving
	\begin{subequations}
		\begin{align}
			\dot E^X &= \dot W^{X \to Y} = - \dot W^{Y \to X} = - \dot E^Y\\
			\dot I^X &= - \dot I^Y\,,
		\end{align}
	\end{subequations}
	\blue{i.e., if one subsystem's dynamics increase the average energy or mutual information, the dynamics of the other must compensate this change accordingly, to ensure constant energy and mutual information at steady state.}
	These relations are especially useful for the dynamics of biological systems which can often be modelled as at steady state.
	
	\subsection{Marginal and conditional entropy productions}
	Note that the
	\blue{subsystem-specific}
	entropy productions $\dot\Sigma^X$ and $\dot\Sigma^Y$ are neither marginal nor conditional entropy productions, i.e., they do not result from the time-reversal statistics of the non-Markovian marginal processes obtained by only observing the $X$- or $Y$-dynamics or of the statistics of trajectories of one subsystem conditioned on the trajectory of the other.
	
	It is possible to define such marginal and conditional entropy productions for bipartite Markov processes. As shown by Crooks and Still~\cite{Crooks2019_Marginal}, the total entropy production $\dot\Sigma$ is then split into nonnegative marginal and conditional contributions. Similarly to the
	\blue{subsystem-specific}
	entropy production rates in (\ref{eq:local_EP_x_info_flow},\ref{eq:local_EP_y_info_flow}), which contain the information flows $\dot I^X$ and $\dot I^Y$, the resulting marginal and conditional entropy productions contain information-theoretic exchange terms. Unlike the information-flow formalism presented here, such a splitting is not symmetric: this may be natural when there is a clear distinction between the subsystems, e.g., in the context of a sensor influenced by an external environment signal (section~\ref{sec:sensor_setups}), but perhaps less so when one has reason to treat the subsystems on equal footing.

	\subsection{Tighter second laws and information engines} \label{sec:tighter_second_laws}
	Historically, the question of how to incorporate information into a thermodynamic theory so as to restore the second law's validity has attracted much interest. Discussions ranged around the thought experiment of \emph{Maxwell's demon}~\cite{Maxwell1867_Life,Leff2003_Maxwells}, with well-known contributions from Szilard~\cite{Szilard1929}, Landauer~\cite{Landauer1961_Irreversibility}, and Bennett~\cite{Bennett1982_The_thermodynamics}. Within stochastic thermodynamics, Maxwell's demon has been formalized as a process with (repeated) feedback~\cite{Cao2009_Thermodynamics,Sagawa2010_Generalized,Ponmurugan2010_Generalized,Horowitz2010_Nonequilirbium,Sagawa2012_Nonequilibrium} and interactions with an \emph{information reservoir} (often modeled as a \emph{tape of bits})~\cite{Mandal2012_solvable_model,Mandal2013_Refrigerator,Barato2014_Unifying,Barato2014_Information_Reservoirs}.
	
	The advent of increasingly refined experimental techniques for microscale manipulation has enhanced the prospect of finding realizations of Maxwell's thought experiment in real-world molecular machinery, stimulating a formalization of the \emph{thermodynamics of information}~\cite{Parrondo2015_Thermodynamics}. The bulk of the experimental realizations demonstrating the possibility of information engines utilize some kind of time-dependent external control~\cite{Toyabe2010_Experimental,Camati2016_Experimental,Cottet2017_Observing,Masuyama2018_Information-to-work,Koski2014_Experimental_Realiz,Chida2017_Power,Admon2018_Experimental,Paneru2018_Optimal,Ribezzi2019_Large,Paneru2020_Efficiency}. In a recent example, an optically trapped colloidal particle $X$ is ratcheted against gravity without the trap $Y$ transducing any work $W^{Y\to X}$ to it, thus enabling the complete conversion of heat to actual mechanical output work $-W^X_\mathrm{mech}$ in the gravitational potential~\cite{Saha2021_Maximizing,Lucero2021_Maximal}.
	
	The picture of autonomous interacting subsystems does not naturally allow such a clear distinction between measurement and feedback, or between system and tape~\cite{Shiraishi2015_Role}. Instead, \emph{continuous Maxwell demons} are identified by current reversals, apparently making heat flow against the direction indicated by the second law~\cite{Strasberg2013_Physical_model_Maxwell_demon,Koski2015_On-chip,Ciliberto2020_Autonomous_demon,Freitas2021_Characterizing}. In this context the information-flow formalism produces a bound on apparent second-law violations in one subsystem using an information-theoretic quantity.
	
	We are now in a position to assess the role of information flows in the operation of two-component systems and make contact with Maxwell's demon. We focus on the
	\blue{specific}
	second law applied to the $X$-subsystem. Rearranging~\eqref{eq:local_EP_x_cond_entropy}, we obtain:
	\begin{equation}
		\blue{\dot S[X|Y]}
		- \frac{\dot Q^X}{k_\mathrm{B}T} \geq - \dot I^Y \,, \label{eq:local_2nd_law_x}
	\end{equation}
	where the LHS is a conventional expression for the entropy production due to system $X$ (entropy change of the system state $X$ at fixed $Y$, minus heat flow $\dot Q^X$ into the system) and the RHS is an information-theoretic quantity measuring an aspect of correlation between $X$ and another system $Y$. 
	
	Let us distinguish two cases: (1) If $\dot I^Y < 0$---naturally arising whenever there is no feedback from $X$ to $Y$---\eqref{eq:local_2nd_law_x} represents an \emph{improved lower bound} on the traditional expression for entropy production. 
	
	(2) If $\dot I^Y > 0$, \eqref{eq:local_2nd_law_x} states that the traditional expression for entropy production can become negative, in apparent contradiction to the second law. This can reasonably be interpreted as a Maxwell-demon setup, and in this continuous-time formalism can immediately be applied to autonomous Maxwell demons such as~\cite{Strasberg2013_Physical_model_Maxwell_demon,Koski2015_On-chip,Ciliberto2020_Autonomous_demon,Freitas2021_Characterizing}.
	
	In the following two sections~\ref{sec:sensor_setups} and \ref{sec:engine_setups} we discuss both cases in detail.

	\section{Sensors: external Y-dynamics}
	\label{sec:sensor_setups}
	The performance limits of biomolecular sensors such as those found in \emph{Escherichia coli} have gained attention
	\blue{~\cite{Bialek2005_Physical,Tu2007_Nonequilibrium,Lan2016_Information,Mattingly2021_Echerichia}.}
	As observed by Berg and Purcell~\cite{Berg1977_Physics}, the main challenge faced by sensors tasked with measuring concentrations in the microscopic world is the stochastic nature of their input signal, i.e., the irregular arrival and binding of diffusing ligands; different strategies can improve inference of ligand concentration~\cite{Endres2009_Maximum,Govern2012_Fundamental,ten_Wolde2016_Fundamental}.
	
	Sensing has also been studied from an information-thermodynamics perspective, where the main question revolves around the minimum thermodynamic cost to achieve a given sensor accuracy. Maintaining correlation between an internal downstream signalling network and an external varying environment is costly~\cite{Lan2012_Energy,Mehta2012_Energetic,Govern2014_Optimal,Sartori2014_Thermodynamic,Govern2014_Energy,Bo2015_Thermodynamic} and involves erasing and rewriting a memory, analogous to a Maxwell demon~\cite{Ouldridge2017_Thermodynamics}. Here, we focus on a high-level characterization of biomolecular sensing that uses bipartite Markov processes.
	
	Specifically, in a \emph{sensor setup}, the stochastic dynamics of one of the subsystems (the environmental signal) are independent of the other (the sensor). Figure~\ref{fig:paradigmatic_examples}(B) shows an example of a sensor setup inspired by the signaling network involved in \emph{E.~coli} chemotaxis. Let $Y$ be an external process (e.g., whether a ligand is bound to the receptor) that influences the transition rates of the sensor $X$, but whose transition rates are independent of $X$: $R^x_{yy'} = R_{yy'}$. This implies the nonpositivity of the $Y$-information flow in~\eqref{eq:info_flow_y_Markov}:
	\begin{subequations}
		\begin{align}
			\dot I^Y &
			\blue{=}
			\sum_{x,y>y'} R_{yy'}\, p(x,y')\,\ln\frac{p(x|y)}{p(x|y')}\nonumber\\
			&\qquad- \sum_{x,y>y'} R_{y'y}\, p(x,y) \ln\frac{p(x|y)}{p(x|y')} \label{eq:Iy_sensor_derivation_1}\\
			&
			\blue{=}
			\sum_{x,y'>y} R_{y'y}\, p(x,y)\,\ln\frac{p(x|y')}{p(x|y)}\nonumber\\
			&\qquad- \sum_{x,y>y'} R_{y'y}\, p(x,y) \ln\frac{p(x|y)}{p(x|y')} \label{eq:Iy_sensor_derivation_2}\\ 
			&= -\sum_{y \neq y'} R_{y'y}\, p(y) \underbrace{\sum_x p(x|y) \ln\frac{p(x|y)}{p(x|y')}}_{\geq 0} \label{eq:Iy_sensor_derivation_3}\\
			&\leq 0\,, \label{eq:Iy_sensor_derivation_end}
		\end{align}
	\end{subequations}
	where in \eqref{eq:Iy_sensor_derivation_2} we swapped summation indices $y \leftrightarrow y'$ in the first sum, and in \eqref{eq:Iy_sensor_derivation_3} the term with an underbrace is a relative entropy and hence is nonnegative~\cite[Chap. 2.3]{Cover2006_Elements}.
	
	Equation~\eqref{eq:local_EP_x_cond_entropy} thus implies a stronger second-law inequality:
	\begin{align}
		\blue{\dot S[X|Y]}
		- \frac{\dot Q^X}{k_\mathrm{B}T} \geq -\dot I^Y \geq 0\,. \label{eq:sensor_tigher_second_law}
	\end{align}
	The LHS represents the sensor's entropy production, which is lower-bounded by an information-theoretic quantity that has various interpretations in the literature. In the following we will build intuition about this quantity and comment on its relation to the sensor's measuring performance.

	\subsection{Nostalgia and learning rate}\label{sec:Nostalgia}
	The first inequality in~\eqref{eq:sensor_tigher_second_law} was originally pointed out by Still, \emph{et al.} in a discrete-time formalism~\cite{Still2012_Prediction} and for possibly non-Markovian external processes. In that formalism, $-\dot I^Y$ is interpreted as \emph{nostalgia} quantifying the share of the mutual information between $X$ and $Y$ that is not predictive of the immediate future of $Y$ or, equivalently, the rate at which information between $X$ and $Y$ becomes irrelevant due to $Y$ dynamics. A sensor that predicts the future signal worse---in the sense of storing more information that is useless for predicting the next signal state---thus produces more entropy than one that is more predictive, raising the possibility that evolution selects for sensors that make parsimonious predictions.
	
	A second related quantity is the \emph{learning rate} $\ell_\mathrm{x}$ introduced by Barato \emph{et al.}~\cite{Barato2014_Efficiency}. Originally defined for systems in steady state, it is exactly the information flow $\dot I^X$ in \eqref{eq:info_flow_x_Markov}:
	\begin{align}
		\ell_\mathrm{x} &= \dot I^X\,.
	\end{align}
	The learning rate quantifies how the uncertainty in an external signal $Y$ is reduced by the dynamics of $X$, i.e., how much $X$ learns about $Y$:
	\begin{subequations}
		\begin{align}
			\ell_\mathrm{x} &= \sum_{x>x',y} \left[ R^{xx'}_y p(x',y) - R^{x'x}_y p(x,y) \right] \ln\frac{p(y|x)}{p(y|x')}\\
			&= \sum_{x>x',y>y'} \left[ R^{xx'}_{yy'} p(x',y) - R^{x'x}_{y'y} p(x,y) \right] \ln\frac{p(y|x)}{p(y'|x')}\nonumber\\
			&\qquad- \sum_{x,y>y'} \left[ R^{x}_{yy'} p(x,y') - R^{x}_{y'y} p(x,y) \right] \ln\frac{p(y|x)}{p(y'|x)}\\
			&= 
			\blue{-\dot S[Y|X]}
			+ \dot S^Y[Y|X]\,.
		\end{align}
	\end{subequations}
	We used the bipartite assumption~\eqref{eq:bipartite_assumption} in the second line. Here $\dot S^Y[Y|X]$ is the rate of change of $S[Y|X]$ that is due to the $Y$-dynamics~\cite{Hartich2014_Transfer_entropy,Barato2014_Efficiency}.
	
	In the special case of a steady state
	\blue{($\dot S[Y|X] = 0 $),}
	the information flows cancel ($\ell_x = \dot I^X = - \dot I^Y$) and inequality~\eqref{eq:sensor_tigher_second_law} reduces to
	\begin{align}
		-\frac{\dot Q^X}{k_\mathrm B T} \geq \ell_\mathrm{x}\,.
	\end{align}
	This motivated Barato \emph{et al}.\ to define an \emph{informational efficiency}~\cite{Barato2014_Efficiency}, $\eta := -k_\mathrm B T \, \ell_\mathrm{x}/\dot Q^X$, measuring the share of a sensor's dissipation that is used to actually track the environmental signal.
	
	The following series of (in-)equalities sums up the relations between the different measures of information flow:
	\begin{subequations}
		\begin{align}
			\dot I^X &= \underbrace{\ell_\mathrm{x}}_{\mathrm{learning}\, \mathrm{rate}}\\
			&= \lim\limits_{\mathrm{d}t \to 0}\frac{I[X_{t+\mathrm{d}t};X_t] - I[X_{t};Y_t]}{\mathrm{d}t} \qquad\;\; [\eqref{eq:info_flow_change_mut_info_x}]\\
			&= 
			\blue{\quad \dot I}
			- \dot I^Y \qquad\qquad\qquad\qquad\qquad\quad [\eqref{eq:sum_info_flows}]\\
			&= \underbrace{- \dot I^Y}_{\mathrm{(rate}\, \mathrm{of)}\,\mathrm{nostalgia}} \qquad\qquad\qquad [\mathrm{steady}\;\mathrm{state}]\\
			&\geq 0\,. \qquad\qquad\qquad\quad [\mathrm{external}\;Y\mathrm{-dynamics}]
		\end{align}
	\end{subequations}

	\subsection{Other information-theoretic measures of sensor performance}
	While information flow bounds sensor dissipation and has intuitive interpretations in terms of predictive power~\cite{Still2012_Prediction} and learning rate~\cite{Barato2014_Efficiency}, other information-theoretic quantities seem more natural to measure sensor performance.
	
	For example, Tostevin and ten Wolde~\cite{Tostevin2009_Mutual} have calculated the rate of mutual information between a sensor's input and output; however, Barato \emph{et al.}~\cite{Barato2013_Information-theoretic} have shown that this rate is not bounded by the thermodynamic entropy production rate. (The desired inequality requires both the time-forward trajectory mutual information rate and its time-reversed counterpart~\cite{Diana2014_Mutual}.)
	
	Another commonly used quantity to infer causation is the (rate of) \emph{transfer entropy}~\cite{Schreiber2000_Measuring}, which in turn is a version of \emph{directed information}~\cite{Marko1973_The_Bidirectional,Massey1990_Causality} (for a gentle introduction see, e.g.~\cite[section~15.2.2]{Bechhoefer2021_Control}). Much like information flow, this rate also bounds the sensor's entropy production rate~\cite{Hartich2014_Transfer_entropy}; however, in general, it represents a looser bound than the information flow. The transfer-entropy rate measures the growth rate of mutual information between the current environmental signal and the \emph{sensor's past trajectory}. This motivated Hartich \emph{et al.}~\cite{Hartich2016_Sensory} to define \emph{sensory capacity} as the ratio of learning rate and transfer-entropy rate, measuring the share of total information between environmental signal and the entire sensor's past that the sensor's instantaneous state carries. It is maximal if the sensor is an optimal Bayesian filter~\cite{Horowitz2014_Second-law_like,Sarkka2013_Bayesian}.
	
	Finally, a natural quantity to measure a sensor's performance is the \emph{static mutual information} between its state and the environmental signal. Brittain \emph{et al.}~\cite{Brittain2017_What_we_learn} have shown that in simple setups, learning rate and mutual information change in qualitatively similar ways when system parameters are varied; however, in more complex setups with structured environmental processes or feedback from the sensor to the environment, maximizing the learning rate might produce a suboptimal sensor. They rationalize this result by noting that the rate at which the sensor must obtain new information to maintain a given level of static mutual information does not necessarily coincide with the magnitude of that static mutual information.

	\section{Engine setups: feedback from X to Y}
	\label{sec:engine_setups}
	Here, we consider the more general case of an \emph{engine setup} in which the two components $X$ and $Y$ cannot be qualitatively distinguished as an external and an internal process; instead, both components $X$ and $Y$ form a joint system. On a formal level, there now is feedback from $X$ to $Y$, such that~\eqref{eq:Iy_sensor_derivation_end} no longer holds in general and it is not possible to make model-independent statements about the direction of information flow. To make contact with analyses of multi-component molecular machines, we present a few conceptual differences between \emph{external control} by an experimenter and what we call \emph{autonomous control} by another coupled stochastic system.

	\subsection{External vs.\ autonomous control}
	There is a long history of nonequilibrium statistical mechanics motivated by single-molecule experiments. A hallmark of these experiments is dynamical variation by an external apparatus of control parameters such as the position or force of an optical trap~\cite{Bustamante2021_Optical}, magnetic trap~\cite{Megli2009_Single}, or atomic-force microscope~\cite{Neumann2008_Single-molecule}. This \emph{external} control allows an unambiguous identification of work done on a system as the change in internal energy achieved through the variation of control parameters, and heat as the complementary change of internal energy due to the system's dynamics. Sekimoto~\cite{Sekimoto1998_Langevin,Sekimoto2010_Stochastic_Energetics} has identified heat and work for diffusive dynamics described by a Langevin equation; this identification readily carries over to discrete dynamics~\cite{VandenBroeck2015_Ensemble,Seifert2012_Stochastic} and even Hamiltonian dynamics~\cite{Jarzynski2011_Equalities}.
	
	The notion of a deterministic control-parameter trajectory allows, e.g., the derivation of fluctuation theorems~\cite{Jarzynski1997_Nonequilibrium,Crooks1999_Entropy,Seifert2005_Entropy} and the study of how to optimize such a trajectory to minimize the average work done on the system~\cite{Schmiedl2007_Optimal,Then2008_Computing,Sivak2012_Thermodynamic,Zulkowski2012_Geometry,Martinez2016_Engineered,Tafoya2019_Using} or its fluctuations~\cite{Solon2018_Phase_Transition,Blaber2020_Skewed}. Feedback can also be included in the analysis by considering measurements and subsequent modifications to the control-parameter trajectory that depend on measurement outcome~\cite{Cao2009_Thermodynamics,Sagawa2010_Generalized,Ponmurugan2010_Generalized,Horowitz2010_Nonequilirbium,Sagawa2012_Nonequilibrium}.
	
	However, in biological systems, there is generally no dynamical variation of external control parameters. Instead, these systems are autonomous, and stochastic thermodynamics occurs in the context of relatively constant but out-of-equilibrium ``boundary conditions'': a single temperature and a variety of chemical potentials that are mutually inconsistent with a single equilibrium system distribution, thus leading to free-energy transduction~\cite{Brown2020_Theory} when the coupling is sufficiently strong such that not all currents 
	\blue{flow in the direction of their driving force.}
	Increasingly, researchers are modeling molecular machines as multi-component systems with internal flows of energy and information. Examples are the molecular motor $\mathrm{F}_\mathrm{o}\!-\!\mathrm{F}_1$ ATP synthase~\cite{Boyer1997_ATP_Synthase,Yoshida_2001_ATP_synthase, Junge2015_ATP_Synthase} that can be modeled using two strongly coupled subsystems~\cite{Xing2005_Making_ATP,Golubeva2012_Efficiency,Ai2017_Torque-coupled,Fogedby2017_Minimal_model,Sune2019_Efficiency,Lathouwers2020_Nonequilibrium}, or molecular motor-cargo collective systems where sometimes hundreds of motors (such as kinesin, dynein~\cite{Encalada2011_Stable}, and myosin~\cite{Cooke1997_Actomyosin}) work in concert~\cite{Leopold1992_Association,Rastogi2016_Maximum}, leading to different performance trade-offs~\cite{Klumpp2005_Cooperative,Bhat2016_Transport,Bhat2017_Stall,Wagoner2021_Evolution,Leighton2022_Performance,Leighton2022_Dynamic}.
	
	Nonetheless, multi-component systems can be interpreted \emph{as if} the dynamics of one component provide a variation of external control parameters to the other. In this context, it can be useful to identify an upstream (more strongly driven by nonequilibrium boundary conditions) system $Y$ and a downstream (more strongly driven by the coupled upstream system than by the nonequilibrium boundary conditions) system $X$, although the identification of these components may sometimes be ambiguous. This type of \emph{autonomous control} differs from external control in two important aspects: (1) It is \emph{stochastic} since the dynamics of the upstream system are itself stochastic; (2) There is \emph{feedback} from the downstream to the upstream system because the upstream system's dynamics obey
	\blue{local}
	detailed balance~\eqref{eq:generalized_detailed_balance}. Both aspects can lead to counterintuitive results when one naively applies stochastic energetics to one subsystem that is strongly coupled to others~\cite{Large2021_Free-energy}.

	\subsection{Conventional and information engines} \label{sec:conventional_info_engines}
	Let us more closely examine two-component engines, e.g., the  $\mathrm{F}_\mathrm{o}\!-\!\mathrm{F}_1$ ATP synthase sketched in figure~\ref{fig:paradigmatic_examples}(A). Such a molecular machine can be regarded as a kind of \emph{chemical-work transducer} using a stronger upstream chemical gradient to drive a downstream chemical reaction against its natural direction~\cite{Amano2022_Insights,Wachtel2022_Free-Energy}. Recently, the coupling characteristics and energy flows in such systems have received attention~\cite{Sune2019_Efficiency,Lathouwers2020_Nonequilibrium,Large2021_Free-energy,Lathouwers2022_Internal}. 
	
\begin{figure*}[ht]
		\centering
		\includegraphics[width=0.34\linewidth]{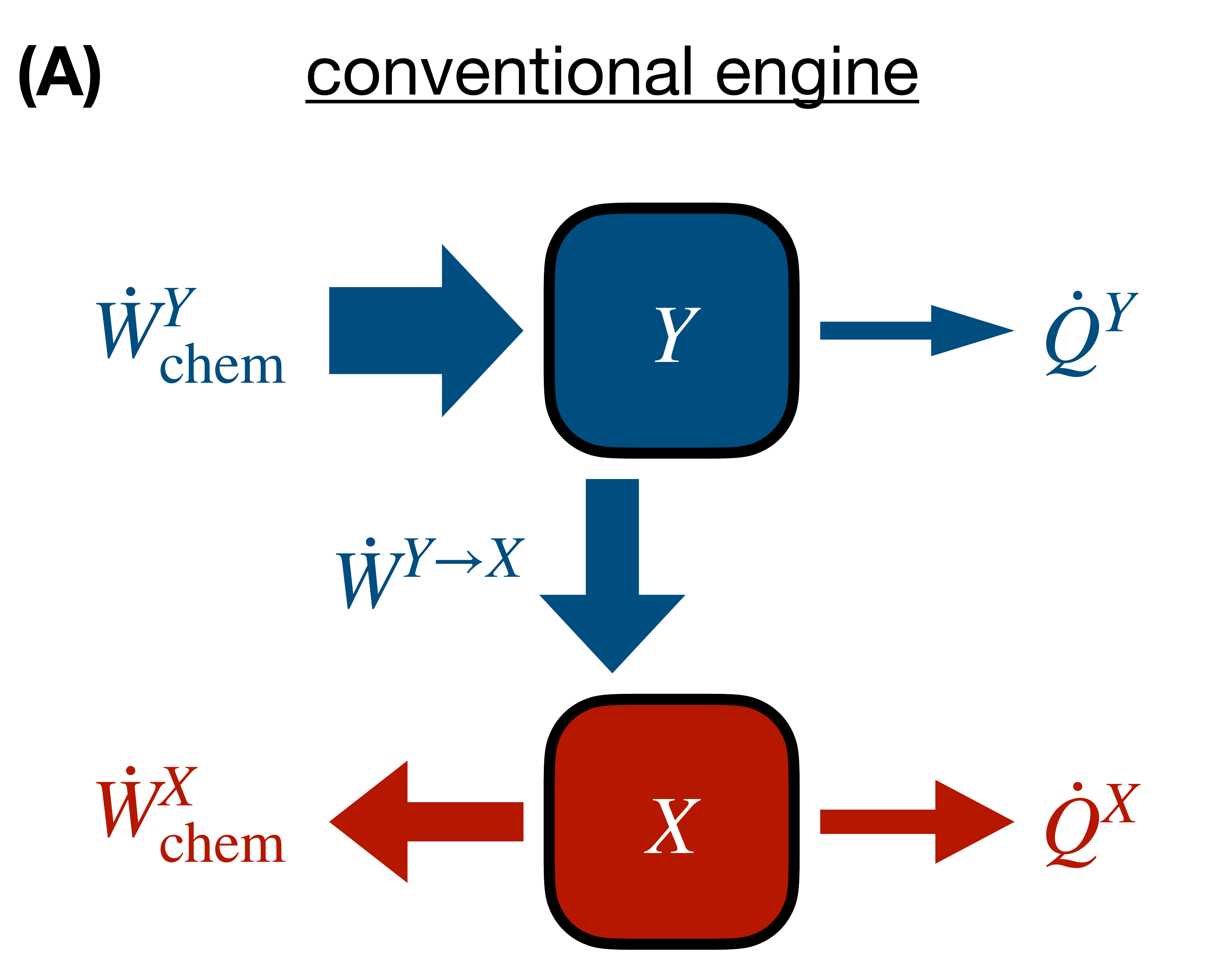}%
		\includegraphics[width=0.34\linewidth]{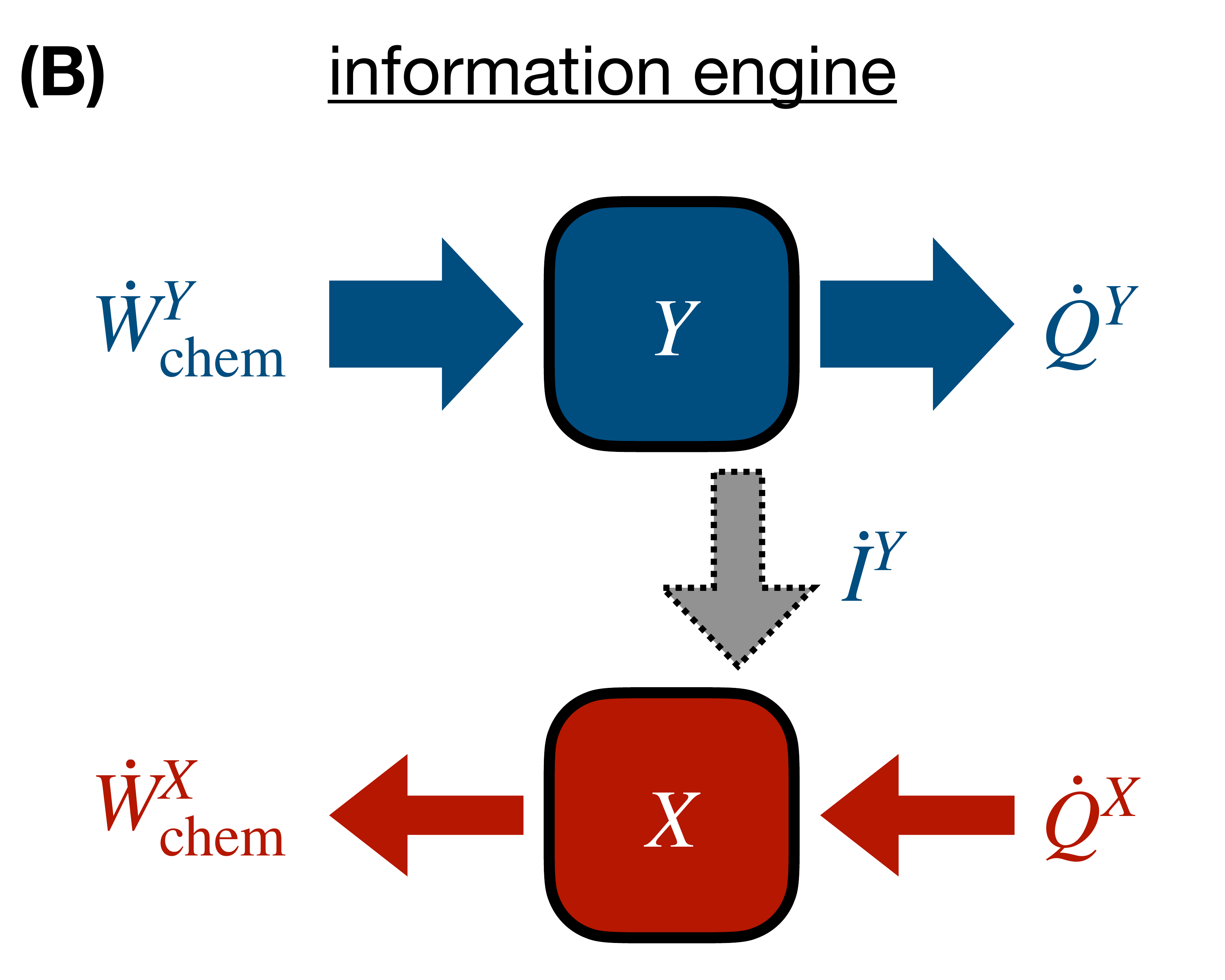}%
		\includegraphics[width=0.34\linewidth]{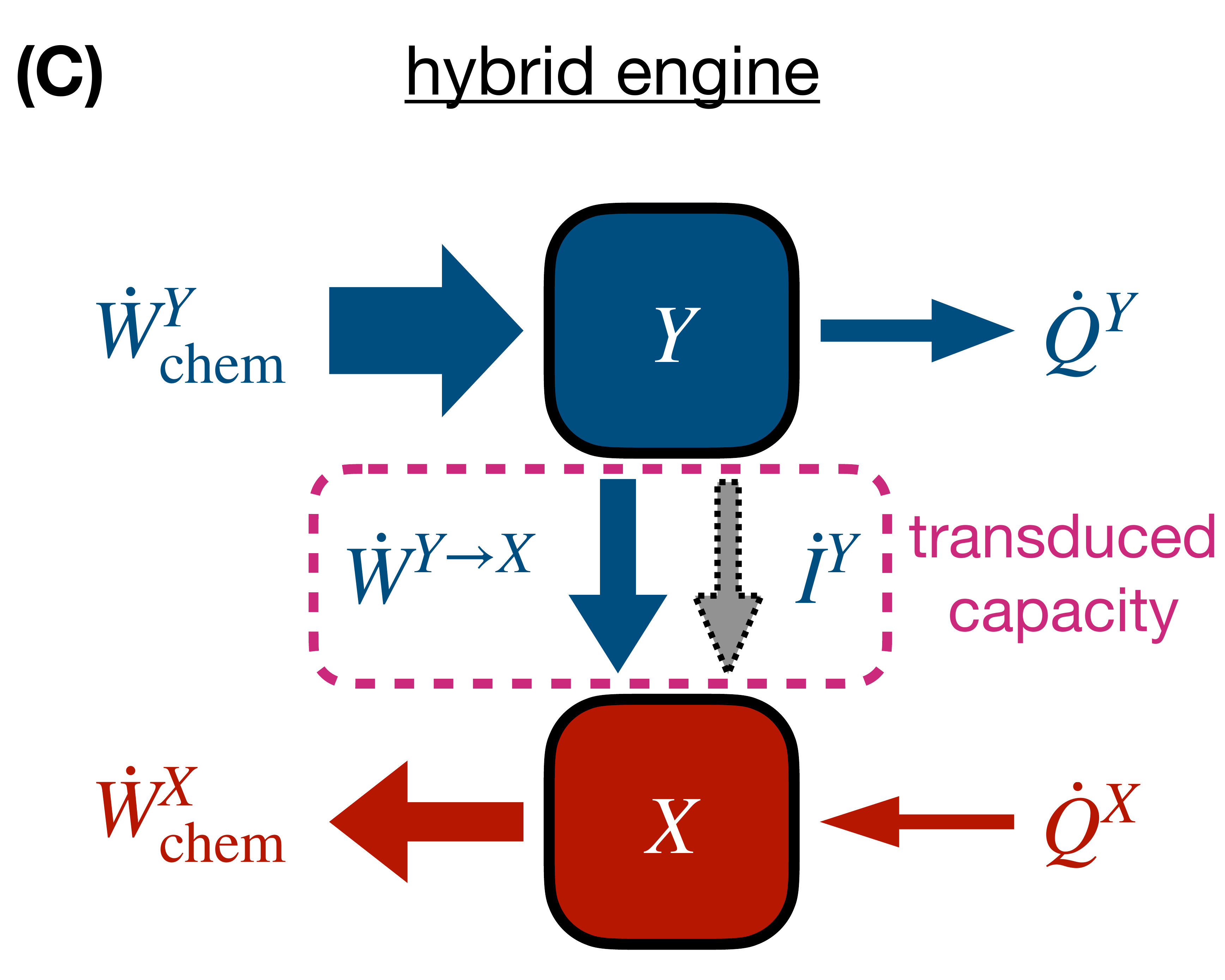}%
		\caption{Different operational modes of a two-component engine converting chemical input power $\dot W^Y_\mathrm{chem}$ to chemical output power $-\dot W^X_\mathrm{chem}$. Arrow direction and thickness respectively indicate the net energy flow's direction and magnitude. \textbf{(A)} In a \emph{conventional engine}, input power $\dot W^{Y \to X}$ is transduced from the upstream component to the downstream component to end up as output power, with heat losses $-\dot Q^Y$ and $-\dot Q^X$ in the process. \textbf{(B)} An \emph{information engine} uses the input power into the upstream component to rectify thermal fluctuations in the downstream component into output power. To achieve this, the upstream component must create information flow $\dot I^Y$ that the downstream component can exploit. \textbf{(C)} A \emph{hybrid engine} uses a mixture of both operation modes. Transduced power and information flow sum to give the transduced capacity.} 
		\label{fig:conventional_info_engine}
	\end{figure*}

	It is natural to consider direct energy flows from an input (chemical) reservoir (e.g., $\dot W^Y_\mathrm{chem}$) through the transduced work (e.g., $\dot W^{Y\to X}$) between subsystems to an output reservoir (e.g., $-\dot W^X_\mathrm{chem}$), with two intermediate heat losses ($-\dot Q^Y$ and $-\dot Q^X$), see figure~\ref{fig:conventional_info_engine}(A). However, a completely different mode of operation is also possible where the input work is not used to transduce energy from $Y$ to $X$ but to rectify thermal fluctuations of $X$, hence converting input heat $\dot Q^X$ into output work $-\dot W^X_\mathrm{chem}$, see figure~\ref{fig:conventional_info_engine}(B). The second setup can be interpreted as an \emph{information engine}, a realization of a Maxwell demon~\cite{Leff2003_Maxwells}, where $X$ is the thermodynamic system controlled by the demon $Y$. Focusing on the energy flows into and out of system $X$ alone would lead an observer to the erroneous conclusion that heat is entirely converted into useful work, a process forbidden by the second law. However this apparent second-law violation results from neglecting the other part of the machine ($Y$), which, to restore the second law, must dissipate more heat into the environment than $X$ converts into work. As discussed in section~\ref{sec:entropy_balance}, the bipartite assumption gives the information flow as a measure to assess the extent to which a given system acts as an information engine.
	
	\subsection{Steady-state transduced capacity} \label{sec:transduced_capacity}
	Our discussion indicates that conventional and information engines can be treated with a common framework, as in~\cite{Amano2022_Insights} where a synthetic molecular motor was analyzed, identifying distinct flows of information and energy with which the upstream subsystem drives the downstream subsystem. For concreteness, let $Y$ be the upstream and $X$ be the downstream subsystem. As shown in~\cite{Barato2017_Thermodynamic_cost,Lathouwers2022_Internal}, combining the subsystem-specific second laws at steady state leads to a simultaneous bound on input and output power in terms of an intermediate quantity, called \emph{transduced capacity} in \cite{Lathouwers2022_Internal}. Substituting the steady-state identities 
	\blue{$\dot S[X] = 0 = \dot S[Y]$}
	and 
	\blue{$\dot I^X = \dot I - \dot I^Y = - \dot I^Y $}
	into the
	\blue{subsystem-specific}
	entropy productions \eqref{eq:local_EP_x_info_flow} and \eqref{eq:local_EP_y_info_flow} gives
	\begin{subequations}
		\begin{align}
			- \frac{\dot Q^X}{k_\mathrm{B} T} + \dot I^Y &\geq 0\\
			- \frac{\dot Q^Y}{k_\mathrm{B} T} - \dot I^Y &\geq 0\,.
		\end{align}
	\end{subequations}
	Using the
	\blue{subsystem-specific}
	first laws in \eqref{eq:local_first_law_x} and \eqref{eq:local_first_law_y} and identifying the transduced power in \eqref{eq:trans_work_y_to_x} gives $\dot E^Y = - \dot E^X = \dot W^{Y \to X}$, so that
	\begin{equation}
		\underbrace{\frac{\dot W^Y_\mathrm{chem} + \dot W^Y_\mathrm{mech}}{k_\mathrm{B} T}}_{\mathrm{input}\;\mathrm{power}} \geq\!\! \underbrace{\frac{\dot W^{Y\to X}}{k_\mathrm{B}T} + \dot I^Y}_{\mathrm{transduced \; capacity}} \!\!\geq \underbrace{- \frac{\dot W^X_\mathrm{chem} + \dot W^X_\mathrm{mech}}{k_\mathrm{B} T}}_{\mathrm{output}\;\mathrm{power}}\,.
	\end{equation}
	This relation suggests that the transduced capacity acts as a \emph{bottleneck} for the conversion of input to output power. The capacity of this bottleneck consists of two distinct pathways, a conventional energetic component $\dot W^{Y \to X}$, with which the upstream subsystem \emph{drives} the downstream subsystem by doing work on it, as well as an information-theoretic component $\dot I^Y$, with which the upstream subsystem creates information between the two subsystems that the downstream subsystem can exploit to generate output power. This hybrid setup is illustrated in figure~\ref{fig:conventional_info_engine}(C).
	
	We expect efficient work transducers to come as close as possible to saturating both inequalities to minimize dissipative losses during their operation. It would be interesting to investigate under which circumstances each of the two pathways leads to the most efficient work transducers and whether real-world biomolecular machinery has evolved to preferentially exploit one over the other.
	

	\section{Conclusions, extensions, and outlook}

	\subsection{Summary}
	In this review we focused on the thermally influenced stochastic dynamics of two-component autonomous systems which are commonly found in biological machinery. We assumed that the dynamics are Markovian and bipartite such that only one subsystem changes its state at a time.
	
	We collected results that show how the bipartite assumption enables the first and second laws of thermodynamics to be split into
	\blue{subsystem-specific}
	versions. The
	\blue{subsystem-specific}
	first laws lead to
	energy flows between the
	\blue{individual}
	subsystems and the environment and to the transduced power -- the energy flow between the subsystems. The
	\blue{subsystem-specific}
	second laws reveal information flows as
	\blue{specific}
	entropic quantities that quantify how the dynamics of a single subsystem change the mutual information shared between the subsystems.
	
	Sensors are a setup to which the formalism applies naturally because an external signal influences the stochastic dynamics of the sensor. Within the framework, the sensor's dissipation (the energy flow) is bounded by an information-theoretic quantity (the information flow) measuring aspects of the influence of the environmental signal on the sensor.
	
	Studying strongly coupled molecular machines within this framework reveals that the more conventional transduced power (the energy flow) from one subsystem to the other is accompanied by the less conventional information flow, which can be interpreted as a hallmark of information engines. Both flows are capable of supporting energy transduction through the coupled system such that conventional and information engines can be studied from the same perspective.

	\subsection{More than two subsystems}
	The question naturally arises whether the information-flow framework can be extended to systems with more than two subsystems. For such systems, Horowitz~\cite{Horowitz2015_Multipartite} defined an information flow $\dot I^{X;Z_{-X}}$, i.e., the information flow between $X$ and all other subsystems $Z_{-X}$ that together compose the global system. This flow is then further refined by identifying which other subsystems can directly interact with $X$, and
	\blue{subsystem-specific}
	second laws of the form of~(\ref{eq:local_EP_x},\ref{eq:local_EP_y}) follow.
	
	However, defining unambiguous directed energy flows as transduced work from one subsystem to another remains challenging for more than two subsystems. Recall that in section~\ref{sec:transduced_work} we argued that in a bipartite system the dynamics of one subsystem at a fixed state of the other can be interpreted as a control-parameter variation on the fixed subsystem. Hence, any potential-energy changes can be interpreted as work done on the fixed subsystem by the dynamic evolution of the other subsystem. Applying this logic to multipartite systems still permits definition of how much work one subsystem contributes to changing the global potential energy, but not the explicit flow between two subsystems. 
	
	Working out conditions under which exact transduced energy flows can be resolved would be an interesting extension and could lead to useful insights for multipartite systems such as energy flows in collections of motors transporting cargoes.

	\subsection{Optimizing coupled work transducers}
	In section~\ref{sec:transduced_capacity} we illustrated that the sum of transduced power and information flows acts as a kind of bottleneck for the transduction of work in two-component engines. Optimizing a given two-component work transducer and studying which of the two pathways maximize throughput seems like an interesting extension.
	
	A first step towards this goal was accomplished in \cite{Lathouwers2022_Internal} for a specific model capturing aspects of $\mathrm{F}_\mathrm{o}\!-\!\mathrm{F}_1$ ATP synthase. It was found that both transduced power and information flow are required to maximize output power and that maximal power tends to lead to equal subsystem entropy productions $\dot\Sigma^X$ and $\dot\Sigma^Y$~(\ref{eq:local_EP_x_info_flow},\ref{eq:local_EP_y_info_flow}).

	\subsection{Application to real-world machinery}
	Finally, it would be interesting to see the information-flow formalism applied to real-world machinery. This would involve measuring and modeling the dynamics of two components of a biomolecular system, e.g., both units of $\mathrm{F}_\mathrm{o}\!-\!\mathrm{F}_1$ ATP synthase, instead of only the dynamics of ${\rm F}_1$ as is conventionally done in most single-molecule experiments and theory~\cite{Yasuda2001,Toyabe2010_Nonequilibrium,Toyabe2011_Thermodynamic,Kawaguchi2014_Nonequilibrium,HayashiPRL15}. This can be accomplished, e.g., by observing two components of a biomolecular system and explicitly calculating information flow, possibly revealing the ratchet mechanism of a Maxwell's demon at work. A first step towards this is found in~\cite{Amano2022_Insights} where a synthetic chemical information motor is analyzed: the authors bridge their information-flow analysis to a chemical-reaction analysis
	\blue{~\cite{Penocchio2022_Information}}
	and identify regimes in which energy or information is the dominant driving mechanism. Another recent contribution in this direction is~\cite{Takaki2022_Information} where information flow has been calculated explicitly for dimeric molecular motors.
	\blue{Finally, Freitas and Esposito recently suggested~\cite{Freitas2022_Maxwell} a macroscopic Maxwell demon based on CMOS technology and analyzed the information flow between its components~\cite{Freitas2022_Information}.}

	\section*{Funding}
	This research was supported by grant FQXi-IAF19-02 from the Foundational Questions Institute Fund, a donor-advised fund of the Silicon Valley Community Foundation. Additional support was from a Natural Sciences and Engineering Research Council of Canada (NSERC) Discovery Grant (D.A.S.) and a Tier-II Canada Research Chair (D.A.S.).
	
	\section*{Acknowledgments}
	We thank Matthew Leighton and John Bechhoefer (SFU Physics) and Mathis Grelier (Grenoble Physics) for helpful conversations and feedback on the manuscript.
	

	\appendix
	\begin{widetext}
		\section{Appendix: Information flows for bipartite Markovian dynamics}\label{app:info_flows_bipartite_Markov}
		Here, we derive the explicit equations \eqref{eq:info_flow_x_Markov} and \eqref{eq:info_flow_y_Markov} from the definitions of information flows~\eqref{eq:info_flow_change_mut_info_x} and \eqref{eq:info_flow_change_mut_info_y} for bipartite Markov processes.
		
		Consider the joint probability of $X$ and $Y$ at different times:
		\begin{subequations}
			\begin{align}
				p(X_{t+\mathrm{d}t}\!=\!x,y_t) &\approx p(X_{t}\!=\!x,y_t)+ \mathrm{d}t\, \sum_{x'} \left[ R^{xx'}_{y_t} p(X_t\!=\!x',y_t) - R^{x'x}_{y_t} p(X_t\!=\!x,y_t) \right] \label{eq:app_joint_prob_x}\\
				p(x_t,Y_{t+\mathrm{d}t}\!=\!y) &\approx p(x_t,Y_{t}\!=\!y)+ \mathrm{d}t\, \sum_{y'} \left[ R^{x_t}_{yy'} p(x_t,Y_t\!=\!y') - R^{x_t}_{y'y} p(x_t,Y_t\!=\!y) \right]\,, \label{eq:app_joint_prob_y}
			\end{align}
		\end{subequations}
		where we have used the Master equation~\eqref{eq:master_equation} together with the bipartite assumption~\eqref{eq:bipartite_assumption} to expand the probability to first order in $\mathrm{d}t$.
		
		Summing over $y_t$ and $x_t$, respectively, gives the marginal probabilities
		\begin{subequations}
			\begin{align}
				p(X_{t+\mathrm{d}t}\!=\!x) &\approx p(X_{t}\!=\!x) + \mathrm{d}t \sum_{x',y_t} \left[ R^{xx'}_{y_t} p(X_t\!=\!x',y_t) - R^{x'x}_{y_t} p(X_t\!=\!x,y_t) \right] \label{eq:app_marginal_prob_x}\\
				p(Y_{t+\mathrm{d}t}\!=\!y) &\approx p(Y_{t}\!=\!y) + \mathrm{d}t \sum_{x_t,y'} \left[ R^{x_t}_{yy'} p(x_t,Y_t\!=\!y') - R^{x_t}_{y'y} p(x_t,Y_t\!=\!y) \right]\,, \label{eq:app_marginal_prob_y}
			\end{align}
		\end{subequations}
		
		\blue{Inserting the expanded joint~(\ref{eq:app_joint_prob_x},\ref{eq:app_joint_prob_y}) and marginal~(\ref{eq:app_marginal_prob_x},\ref{eq:app_marginal_prob_y}) probabilities into the definition of entropy~\cite{Cover2006_Elements} allows us to expand}
		the conditional entropies for small $\mathrm{d}t$,
		\begin{subequations}
			\begin{align}
				S[Y_t|X_{t+\mathrm{d}t}] &= S[X_{t+\mathrm{d}t},Y_t] - S[X_{t+\mathrm{d}t}]\\
				&\approx S[X_t,Y_t] - S[X_t] - \mathrm{d}t \sum_{x,x',y_t} \left[ R^{xx'}_{y_t} p(X_t\!=\!x',y_t) - R^{x'x}_{y_t} p(X_t\!=\!x,y_t) \right]\ln p(X_t\!=\!x,y_t)\\
				&\qquad+ \mathrm{d}t \sum_{x,x',y_t} \left[ R^{xx'}_{y_t} p(X_t\!=\!x',y_t) - R^{x'x}_{y_t} p(X_t\!=\!x,y_t) \right]\ln p(X_t=x)\,,\nonumber
			\end{align}
		\end{subequations}
		and similarly
		\begin{align}
			S[X_t|Y_{t+\mathrm{d}t}] &\approx S[X_t,Y_t] - S[Y_t]- \mathrm{d}t \sum_{x_t,y,y'} \left[ R^{x_t}_{yy'} p(x_t,Y_t\!=\!y') - R^{x_t}_{y'y} p(x_t,Y_t\!=\!y) \right]\ln p(x_t,Y_t\!=\!y)\\
			&\qquad+ \mathrm{d}t \sum_{x_t,y,y'} \left[ R^{x_t}_{y,y'} p(x_t,Y_t\!=\!y') - R^{x_t}_{y',y} p(x_t,Y_t\!=\!y) \right]\ln p(Y_t=y)\,,\nonumber
		\end{align}
		and substituting into
		\eqref{eq:info_flow_change_mut_info_x} 
		gives
		\begin{subequations}
			\begin{align}
				\dot I^X &:= \lim\limits_{\mathrm{d}t \to 0}\frac{I[X_{t+\mathrm{d}t};Y_t] - I[X_{t};Y_t]}{\mathrm{d}t} = \lim\limits_{\mathrm{d}t \to 0}\frac{S[Y_t|X_{t}] - S[Y_t|X_{t+\mathrm{d}t}]}{\mathrm{d}t}\\
				&= \sum_{x,x',y_t} \left[ R^{xx'}_{y_t} p(X_t\!=\!x',y_t) - R^{x'x}_{y_t} p(X_t\!=\!x,y_t) \right]\ln p(y_t|X_t\!=\!x)\\
				&= \sum_{x>x',y_t} \left[ R^{xx'}_{y_t} p(X_t\!=\!x',y_t) - R^{x'x}_{y_t} p(X_t\!=\!x,y_t) \right] \ln\frac{p(y_t|X_t\!=\!x)}{p(y_t|X_t\!=\!x')}\,,
			\end{align}
		\end{subequations}
		which is \eqref{eq:info_flow_x_Markov}. Similarly, \eqref{eq:info_flow_change_mut_info_y} becomes
		\begin{subequations}
			\begin{align}
				\dot I^Y &:= \lim\limits_{\mathrm{d}t \to 0}\frac{I[X_t;Y_{t+\mathrm{d}t}] - I[X_{t};Y_t]}{\mathrm{d}t} = \lim\limits_{\mathrm{d}t \to 0}\frac{S[X_t|Y_{t}] - S[X_t|Y_{t+\mathrm{d}t}]}{\mathrm{d}t}\\
				&= \sum_{x_t,y,y'} \left[ R^{x_t}_{yy'} p(x_t,Y_t\!=\!y') - R^{x_t}_{y'y} p(x_t,Y_t\!=\!y) \right]\ln p(x_t|Y_t\!=\!y)\\
				&= \sum_{x_t,y>y'} \left[ R^{x_t}_{yy'} p(x_t,Y_t\!=\!y') - R^{x_t}_{y'y} p(x_t,Y_t\!=\!y) \right] \ln\frac{p(x_t|Y_t\!=\!y)}{p(x_t|Y_t\!=\!y')}\,,
			\end{align}
		\end{subequations}
		which is \eqref{eq:info_flow_y_Markov}.
	\end{widetext}
	
	\bibliography{references}
\end{document}